\documentclass{aa}  

\usepackage{graphicx}
\usepackage{txfonts}
\begin{document}

   \title{FSR\,1776: a new globular cluster in the Galactic bulge?}

   \author{B. Dias
          \inst{1}
          \and
          T. Palma\inst{2,3}
          \and
          D. Minniti\inst{4,5}
          \and
          J. G. Fernández-Trincado\inst{6}
          \and
          J. Alonso-Garc{\'i}a\inst{7,8}
          \and
          B. Barbuy\inst{9}
          \and
          J. J. Clariá\inst{2,3}
          \and
          M. Gomez\inst{4}
          \and
          R. K. Saito\inst{10}
          }

   \institute{Instituto de Alta Investigaci\'on, Universidad de Tarapac\'a, Casilla 7D, Arica, Chile\\
   \email{bdiasm@academicos.uta.cl}
    \and
    Universidad Nacional de C\'ordoba, Observatorio Astron\'omico de C\'ordoba, Laprida 854, 5000 C\'ordoba, Argentina
    \and
    Consejo Nacional de Investigaciones Cient\'ificas y T\'ecnicas (CONICET), Godoy Cruz 2290, Ciudad Aut\'onoma de Buenos Aires, Argentina
    \and
    Departamento de Ciencias F\'isicas, Facultad de Ciencias Exactas, Universidad Andr\'es Bello, Av. Fern\'andez Concha 700, Las Condes, Santiago, Chile
    \and
    Vatican Observatory, V00120 Vatican City State, Italy
    \and
    Instituto de Astronom\'ia, Universidad Cat\'olica del Norte, Av.
Angamos 0610, Antofagasta, Chile
    \and
    Centro de Astronomía (CITEVA), Universidad de Antofagasta, Av. Angamos 601, Antofagasta, Chile
    \and
    Millennium Institute of Astrophysics, Nuncio Monse\~nor Sotero Sanz 100, Of. 104, Providencia, Santiago, Chile
    \and
    Universidade de S\~ao Paulo, IAG, Rua do Mat\~ao 1226, Cidade Universit\'aria, S\~ao Paulo 05508-900, Brazil
    \and
    Departamento de Física, Universidade Federal de Santa Catarina, Trindade, 88040-900 Florianópolis, SC, Brazil
    }

   \date{Received XXX; accepted XXX}

 
  \abstract
   { Recent near-IR surveys have uncovered a plethora of new globular cluster (GC) candidates towards the Milky Way bulge. These new candidates need to be confirmed as real GCs and properly characterised.}
   {We investigate the physical nature of FSR\,1776, a very interesting star cluster projected towards the Galactic bulge. This object was originally classified as an intermediate-age open cluster and has recently been re-discovered independently and classified as a GC candidate (Minni 23). Firstly, we aim at confirming its GC nature; secondly we determine its physical parameters.}
   {The confirmation of the cluster existence is checked using the radial velocity (RV) distribution of a MUSE data cube centred at FSR\,1776. The cluster parameters are derived from isochrone fitting to the RV-cleaned colour-magnitude diagrams (CMDs) from visible and near-infrared photometry taken from VVV, 2MASS, DECAPS, and Gaia altogether.}
   {The predicted RV distribution for the FSR\,1776 coordinates, considering only contributions from the bulge and disc field stars, is not enough to explain the observed MUSE RV distribution. The extra population (12\% of the sample) is FSR\,1776 with an average RV of $-103.7\pm 0.4~{\rm km}\,{\rm s}^{-1}$. The CMDs reveal that it is 10$\pm$1~Gyr old and metal-rich population with [Fe/H]$_{phot}\approx + 0.2\pm$0.2, [Fe/H]$_{spec}=~+0.02\pm0.01~(\sigma~=~0.14$~dex), located at the bulge distance of 7.24$\pm$0.5~kpc with A$_{\rm V}$ $\approx$ 1.1~mag. The mean cluster proper motions are ($\langle\mu_{\alpha}\rangle,\langle\mu_{\delta}\rangle$) $=$ ($-2.3\pm1.1,-2.6\pm0.8$) ${\rm mas\, yr^{-1}}$.}
   {FSR\,1776 is an old GC located in the Galactic bulge with a super-solar metallicity, among the highest for a Galactic GC. This is consistent with predictions for the age-metallicity relation of the bulge, being FSR\,1776 the probable missing link between typical GCs and the metal-rich bulge field. High-resolution spectroscopy of a larger field of view and deeper CMDs are now required for a full characterisation.}

   \keywords{Galaxy: bulge -- Galaxy: stellar content --
                globular clusters: FSR\,1776 
               }

   \maketitle

%

\section{Introduction}

The Milky Way should host a larger population of globular clusters (GCs), assuming that its GC population correlates well with its total mass \citep[e.g.,][]{harris+13}. The missing GCs could have been dynamically dissolved or could be hidden in the disc and bulge. Both scenarios present challenges to be proved but the second case can now be assessed with deep near-infrared (NIR) photometric surveys. Most of the missing GCs might have lower masses than typical GCs, which makes them intrinsically difficult to find in the midst of a crowded and extinct field. The currently known Galactic GC sample has been proven useful to understand the formation of our Galaxy and has shed light on a diversity of processes, such as proto-galactic collapse, accretion, galaxy collisions, cannibalism, star bursts \citep[e.g.,][]{vandenbergh93,west+04,kruij+12}, among other tumultuous events that might have taken place. The Galactic bulge is probably the most complex region of the Milky Way, and it still is the less studied component due to the high reddening \citep[e.g.][]{barbuy+18}. The missing low-mass GCs, in particular in the Galactic bulge, should be identified in order to have a complete understanding of the Milky Way history.

The high extinction and stellar density in the Galactic plane, due to foreground and background stars along the line of sight, made it very difficult to search for GCs in this region. There have been few optical direct visual inspections or automatic detection algorithms in the Galactic plane. Therefore, new searches have begun to be made using other photometric bands and non direct detection methods. The use of NIR wavelengths has greatly improved the detection of new GC candidates, especially in the inner Galactic bulge region. The interstellar clouds become more transparent in the NIR than in the optical region and the bright red giant (RG) stars appear brighter. Indirect methods to recognise new GC candidates have led us to look for different old population tracers, such as RR Lyrae stars, Type II Cepheids and red clump (RC) stars. An over-density of such tracers detected in a small space region could imply that those stars are physical members of an underlying and highly obscured GC.

During the last decade, struggles have been made, particularly in the Galactic bulge region, to search for new hidden GCs, which resulted in a significant increase in newly detected GC candidates. An updated catalogue of Galactic associations, open and GCs, embedded groups, and all the corresponding newly detected candidates was published by \citet{bica+19}. The Vista Variables in the Via Lactea (VVV) is a deep near-NIR survey carried out between 2010 and 2016. This survey has been  mapping the most reddened and crowded regions of the Milky Way \citep[MW,][]{minniti+10,saito+12}, allowing the discovery of an important number of GC candidates, not only by visual inspection and/or through photometric analyses searching for concentrations of RC stars \citep{minniti+11,monibidin+11,borissova+14} but also by searching over-densities of different types of variable stars, which are  typical tracers of old stellar populations \citep{minniti+17a,minniti+17b,minniti+17c,minniti+19,minniti+20,minniti+21}. So far, a total of 350 globular candidates were listed as the so-called Minni clusters.

\citet{piatti18} analysed the structure and distances of Minni\,01 to
Minni\,22 based on 2MASS data and concluded that all of them are real bulge GCs, that means a 100\% of real clusters. \citet{gran+19} analysed VVV CMDs along with Gaia DR2 and VVV proper motions (PMs) of Minni\,01 to Minni\,84 and concluded that none of these are real GCs, that means a 0\% of real clusters or at least they are below their detection limit, i.e., none of the cluster candidates are high-mass objects and all of them have PMs similar to the bulk of the bulge. Moreover, they estimate that only about 7\% of such cluster candidates should be real. The conclusions by \citet{piatti18} and \citet{gran+19} cannot be both right. \citet{palma+19} analysed cleaned VVV CMDs and Gaia DR2 PMs of Minni\,23 to Minni\,44 and concluded that Minni\,25, 27, 36, 43, and 44 are not real clusters, and that the best candidates are Minni\,23 (=FSR\,1776) and Minni\,28, which leads to a 77\% of real clusters. The question of whether the Minni clusters are real GCs at the low luminosity tail of the GC luminosity function, being the skeletons of disrupted GCs, or else intermediate-age GCs (or a combination of all of these possibilities) is still open. 

In this work we focus on one of the best GC candidates suggested in \citet{palma+19}. FSR\,1776
was discovered by detecting an over-density of RR\,Lyrae stars and red giant stars on sky \citep{minniti+17a}. It is located at $(RA,DEC)_{2000} = (17.904^h, -36.1524^{\circ})$, matching the previously known cluster FSR\,1776 \citep{fsr07,kharch+13,kharch+16}, whose catalogued coordinates are $(RA,DEC)_{2000} = (17.9^h, -36.145^{\circ})$. Now we add a further dimension to the analysis, the radial velocities (RVs) and spectroscopic metallicities provided by MUSE spectra of a large sample of stars around FSR\,1776.
We aim to clarify whether we can (i) confirm, (ii) rule out or (iii) find evidence to further investigate with more observational data if
FSR\,1776 is a low-mass bulge GC.

The paper is organised as follows. Section 2 gives the MUSE observations and Section 3 shows the distribution of RVs. We present in Section 4 the CMDs and the determination of some cluster physical parameters, while cluster orbits are computed in Section 5. An analysis and discussion of the results is presented in Section 6. Section 7 summarises our main conclusions.

%
\section{MUSE observations}

\begin{figure}
\centering
\includegraphics[width=0.9\columnwidth]{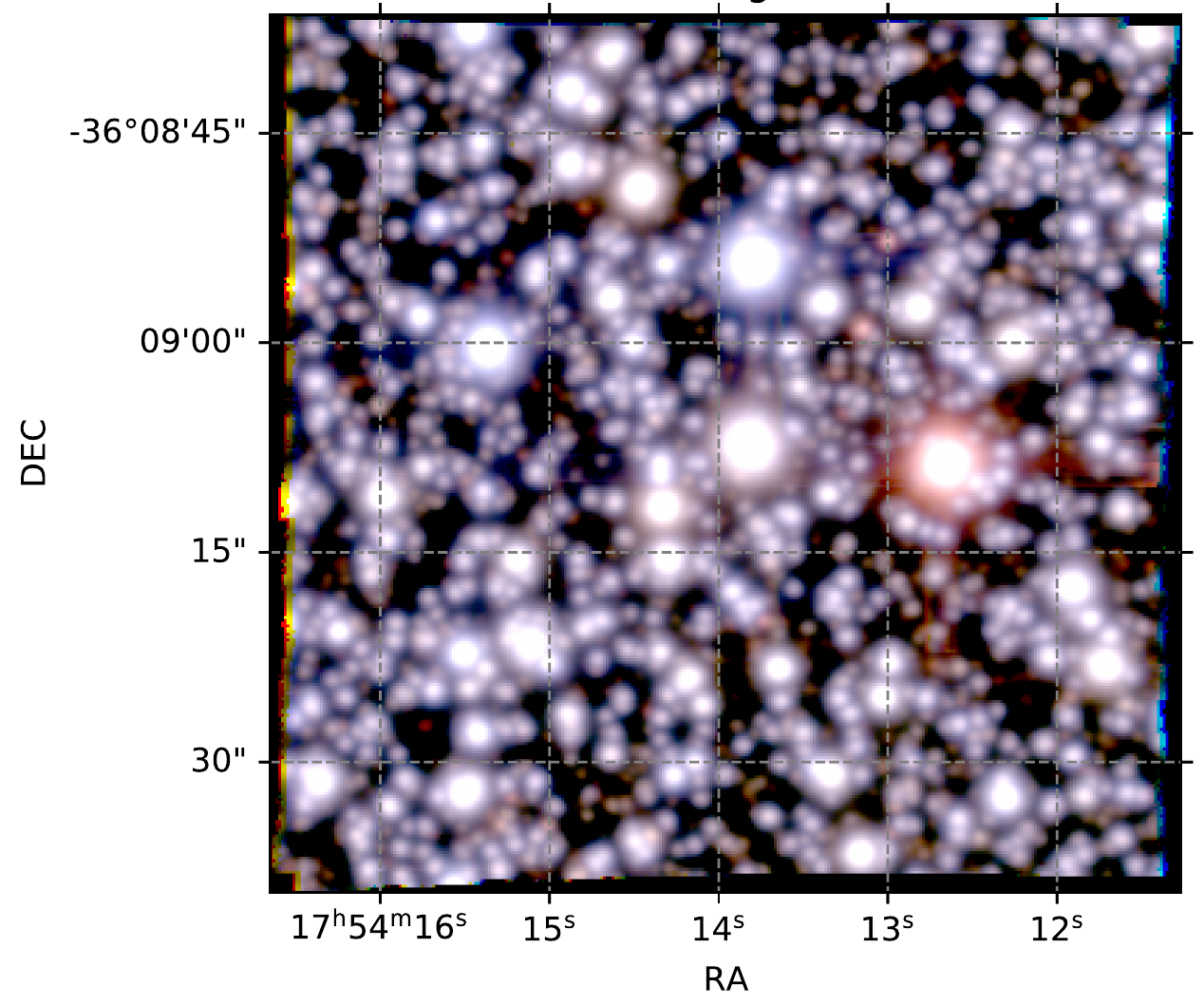}
\caption{Composite coloured image of the MUSE data cube. The original images were taken by convolving the transmission curves of the DECam filters g, r, i with the MUSE cubes. The full field of view of $1\arcmin\times1\arcmin$ is shown. The flux scale is a hyperbolic sine to enhance fainter stars, which are important in the present analysis.
}
\label{fig:image}
\end{figure}

Spectroscopy for a large number of stars is required in order to distinguish field stars and cluster members in the crowded region of
FSR\,1776. Ca\,II triplet (CaT) observed spectra with low-resolution (R$\sim$3,000) gives RV precision of the order of a few ${\rm km\,s^{-1}}$. The estimated velocity dispersion of the bulge and disc at FSR\,1776
coordinates are $\sigma_{\rm RV_{GC}}\sim90{\rm\,km\,s^{-1}}$ \citep{zoccali+14} and $\sigma_{\rm RV_{GC}}\sim40{\rm\,km\,s^{-1}}$ \citep{ness+16}, respectively. It would therefore be feasible to find a sharper peak with dispersion of a few ${\rm km\,s^{-1}}$ corresponding to FSR\,1776. In other words, a low-resolution multi-object or integral field spectrograph is ideal. The best choice for this program was the Multi Unit Spectroscopic Explorer (MUSE) at the 8\,m UT4 at the European Southern Observatory (ESO) \citep{bacon+10}, which also has the adaptive optics module GALACSI \citep{stuik+06} to improve the spatial resolution, diminish the crowding effect and enhance the signal-to-noise ratio (SNR) of each stellar spectrum. The observations log is given in Table \ref{tab:log} and a colour-composite image of the MUSE field-of-view is shown in Fig. \ref{fig:image}. Data reduction was carried out with standard ESO pipeline during ESO Phase III, and the reduced images were obtained from the ESO archive. The image reveals the good spatial resolution for crowded regions and the stability of the point spread function (PSF) throughout the field of view.

The stellar spectra were extracted from the MUSE data cube using the software PampelMUSE\footnote{\url{https://gitlab.gwdg.de/skamann/pampelmuse}} \citep{kamann+13,kamman18}. A reference catalogue of stars is a required input to extract individual stellar spectra of these specific coordinates. There is no deep space-based catalogue of this field to be used, VVV photometry is not as deep as the MUSE photometry, and DECAPS photometry only have extra fainter stars that generate very low SNR spectra from the present MUSE data cube and do not affect the extraction of the brighter stars, being just a granulation on the background sky. The adopted catalogue was produced by running DAOPHOT on a slice of the MUSE data cube. This strategy is a proof-of-concept that can be applied in other similar analysis when   there is no space-based input catalogue available. In a few words, PampelMUSE performs PSF photometry on all slices for each provided coordinate including deblending of stars. It also takes into account the PSF change with wavelength based on a function fitted to the actual data. The final individual spectra are composed by the individual fluxes in each slice. For a detailed description and similar applications, see \cite{kamann+13,husser+16,kamman18}.

\begin{table}[!htb]
\centering
\caption{Log of MUSE observations} \label{tab:log}
\begin{tabular}{lr}
\hline
\hline
\noalign{\smallskip}
 Field ID   &  FSR\,1776  \\
 RA         & $17^h54^m14.0^s$ \\
 DEC       &  $-36^{\circ}09'08.0"$ \\
 FOV & $1'\times1'$ \\
 Obs. date &  2018-06-08\\
 Starting time (UT) & 08:52:09 \\
 Exp. time &  3$\times$1035 sec \\ 
 IQ @ $i$  &  0.74$\arcsec$ (AO) \\
 Prog.ID  & 0101.D-0363(A) \\
 PI  & Minniti \\
\noalign{\smallskip}
\hline
\noalign{\smallskip}
\end{tabular}
\end{table}

%
\section{Radial velocities}
\label{sec:RV}
%
\subsection{Cross-correlation}

\begin{figure}[!htb]
    \centering
    \includegraphics[width=\columnwidth]{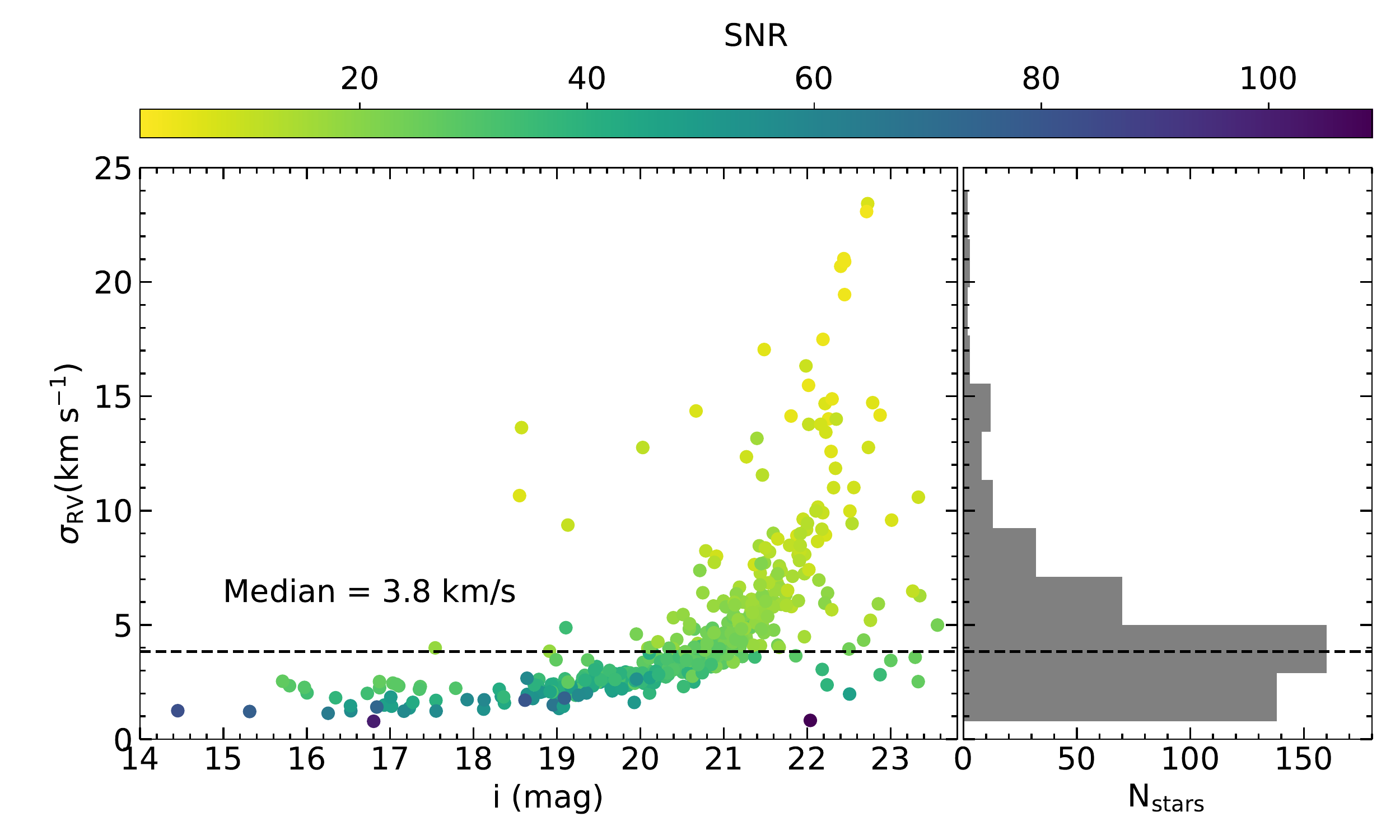}
    \caption{Uncertainties in RV as a function of magnitude for all stars analysed here. The SNR is indicated by the colours, and the histogram shows the median error of $3.8~{\rm km\, s}^{-1}$.}
    \label{fig:rverr}
\end{figure}

Radial velocities were derived with the ETOILE code \citep{katz+11,dias+15} through cross-correlation with a template spectrum from the MILES library \citep{sanchez-blazquez+06} with a similar spectral type for each analysed spectrum, using the visible portion of the spectra bluer than the laser gap (470-575nm). Tests were made with RG stars to compare the derived velocities using also the NIR portion of the spectra in the CaT region and the agreement is very good with a negligible offset of -0.2~${\rm km\, s^{-1}}$ and a dispersion of 2.8~${\rm km\, s^{-1}}$. \cite{valenti+18} derived RVs for individual bulge stars observed with MUSE following a similar procedure to ours and performed Monte Carlo simulations to measure the uncertainties in RVs. We adopted their analytical relations for giant and dwarf stars to calculate the error in RV as a function of SNR and [Fe/H], with SNR given by PampelMUSE and [Fe/H] by ETOILE using full-spectrum fittings (see \citealp{dias+15} and Appendix \ref{app:etoile} for details). We found half of the stars present RV errors smaller than 3.8~${\rm km\, s^{-1}}$ and 66\% of the sample have RV errors smaller than 5.0~${\rm km\, s^{-1}}$, which are of the order or larger than the dispersion in the aforementioned comparison (see Fig. \ref{fig:rverr}). The results were converted to heliocentric RVs using the IRAF\footnote{\url{http://iraf.noao.edu/}} routine {\it rvcorrect}. It is worth noting that a similar analysis was successfully applied before to MUSE stellar spectra of another bulge GC \citep{ernandes+19}.

%
\subsection{RV distribution}

\begin{figure}[!htb]
\centering
\includegraphics[width=\columnwidth]{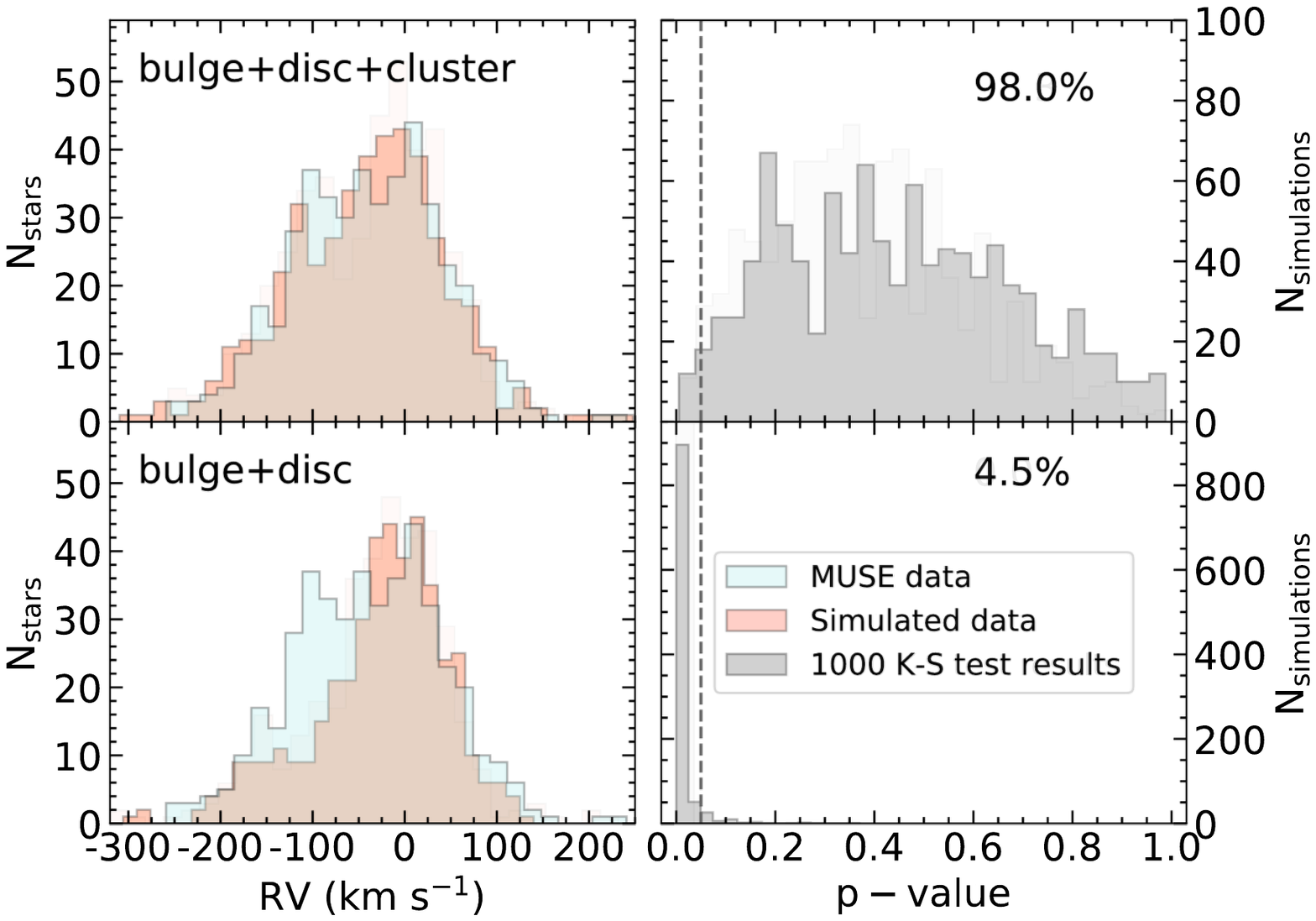}
\caption{Left panels show the MUSE RV distribution compared to an example of simulated RV distribution with and without considering the cluster. Right panels show the distribution of $p-value$ resulting from the comparison of 1000 simulated RV distributions with that from MUSE data. The dashed line indicates $p-value=0.05$ and the percentage refers to the fraction of simulations resulting in $p-value>0.05$.}
\label{fig:minni23RVdist}
\end{figure}

Radial velocities were derived for all 450 extracted spectra from the MUSE data cube and produced the heliocentric RV (RV$_{hel}$) distribution (RVD) of stars within $1\arcmin\times1\arcmin$ around FSR\,1776 displayed in Fig. \ref{fig:minni23RVdist} in turquoise. We compare the observed MUSE RVD with simulated ones to search for a star cluster signature. 

The first step was to generate a RVD containing bulge and disc stars. We used the GIBS survey interpolator \citep{zoccali+14} to estimate the bulge component as RV$_{hel}\sim-50{\rm \,km\,s^{-1}}$ and $\sigma_{\rm RV_{GC}}\sim\,90{\rm \,km\,s^{-1}}$. The disc component was estimated by inspecting Fig.7 of \citet{ness+16} that is a map of RV and $\sigma$ for foreground disc stars from the APOGEE survey. We found RV$_{hel}\sim0{\rm \,km\,s^{-1}}$ and $\sigma_{\rm RV_{GC}}\sim40{\rm \,km\,s^{-1}}$ for the coordinates of FSR\,1776 (see also Appendix \ref{app:gaiaRV} where Gaia RVs corroborate these estimates). The relative fraction of 60\% bulge and 40\% disc stars was estimated running a Besançon model at the same coordinates. Finally, a random sampling of RVD$_{(bulge+disc)}$ following the parameters above was performed to extract 450 stars for a direct comparison with the MUSE RVD. We generated 1000 bootstrap RVDs$_{(bulge+disc)}$ and for each case we calculated the p-value from the Kolmogorov-Smirnov (K-S) comparison between MUSE and simulated RVDs. Only 4.5\% of the simulations yielded $p-value > 0.05$, i.e., it is unlikely that the MUSE RVD is composed only by bulge and disc stars (see bottom panels in Fig. \ref{fig:minni23RVdist}). In fact, the visual inspection of the simulated and observed RVDs shows an excess of stars around RV~$\sim -100~{\rm km\, s^{-1}}$.

\begin{figure*}[!htb]
\centering
\includegraphics[width=0.8\textwidth]{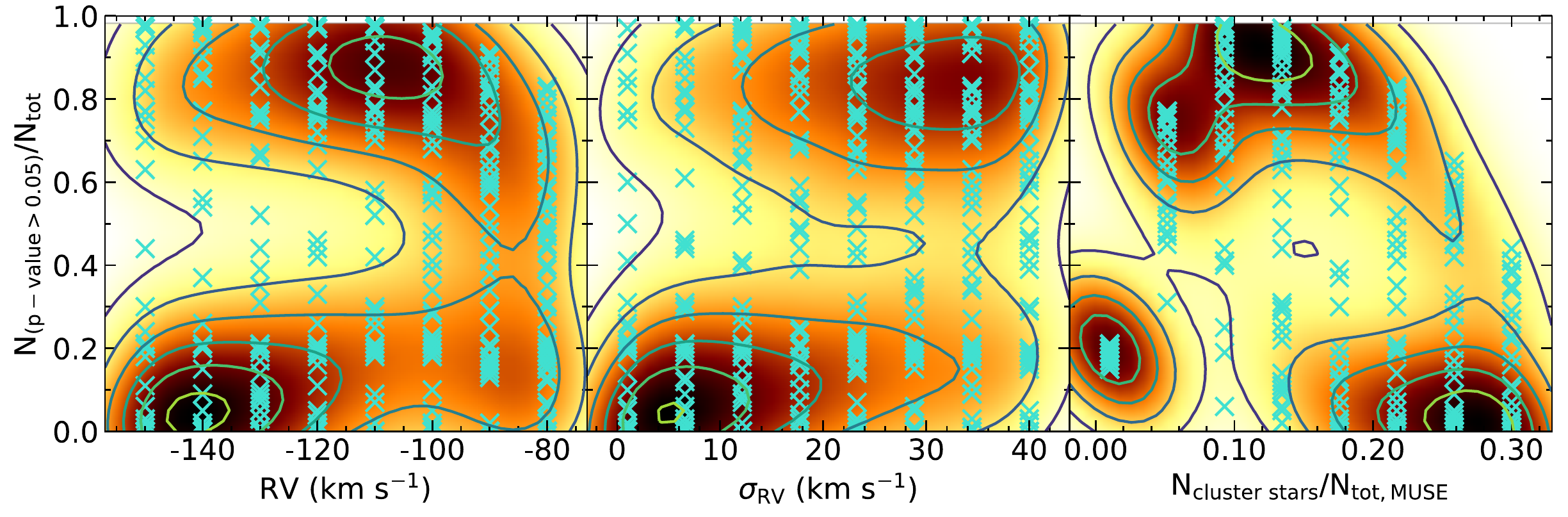}
\caption{Distribution of the fraction of cases with $p-value > 0.05$ versus RV, $\sigma_{RV}$, $f_{clus}$. Turquoise crosses show the result, whereas the orange density regions are the KDE of the points.}
\label{fig:minni23RVfit}
\end{figure*}

The second step was to add a third distribution of stars on the top of RVD$_{(bulge+disc)}$, with a given mean, dispersion, and fraction of total stars, where each of these parameters was allowed to vary in 8 steps within the ranges $-150<$ RV (${\rm km\, s^{-1}}$) $< -80$, $0 < \sigma_{RV} (\rm km\, s^{-1}) < 40$, and $0 < f_{\rm cluster} < 0.3$. These ranges were chosen to give some flexibility for the fit whereas providing information from the excess found above around -100~${\rm km\, s^{-1}}$. For each of the 512 combinations, 100 bootstrap RVD$_{(bulge+disc+cluster)}$ were generated and the fraction of cases with $p-value > 0.05$ from the K-S test was stored. We show the results of all fractions of $p-value > 0.05$ per parameter in Fig. \ref{fig:minni23RVfit}, where a convergence for RV$_{hel}\sim-110{\rm \,km\,s^{-1}}$, $\sigma_{\rm RV_{GC}}\sim32{\rm \,km\,s^{-1}}$ and $f_{cluster}\sim12\%$ becomes evident. For such case, almost all bootstrap RVDs represent well the MUSE RVD. In the upper panels of Fig. \ref{fig:minni23RVdist} we repeat the exercise done for the bottom panels with 1000 bootstrap RVDs including now the third component with the above resulting parameters. In fact, 98\% of the RVDs agree with the MUSE RVD and the visual inspection of simulated and observed RVDs also seems reasonable. 

The most likely explanation for this peculiar third component is the presence of a GC-like structure precisely at the position of FSR\,1776. The velocity dispersion for this third RV component is high compared to typical Milky Way globular clusters (typically within $\sim0.5-18\ {\rm km\ s^{-1}}$, see e.g. \citealp[2010 ed.][]{harris96}). The RV errors do not account for this dispersion (see Fig.\ref{fig:rverr}). It is worth noting that by selecting only the MUSE stars with $\sigma_{RV} < 5{\rm \,km\,s^{-1}}$ (296 stars out of 450) and cleaning outliers outside $-300<RV({\rm \,km\,s^{-1}})<300$, the results are still similar, with RV$_{hel}\sim-95{\rm \,km\,s^{-1}}$ and $\sigma_{\rm RV_{GC}}\sim32{\rm \,km\, s^{-1}}$ and a fraction of $f_{cluster}\sim10\%$, in 94\% of cases with $p-value > 0.05$. The relatively large velocity dispersion would mean that FSR\,1776 is one of the most massive Galactic globular clusters, if in virial equilibrium. If this was the case, we should clearly spot it in the VVV images, which is not the case. Another bulge globular cluster, Terzan\,9, was recently analysed by \citet{ernandes+19} with similar MUSE observations and analysis as done here. They reported a velocity dispersion of $\sim22{\rm \,km\,s^{-1}}$ that is a factor $3\times$ larger than the result reported in the compilation by Baumgardt et al.\footnote{ \url{https://people.smp.uq.edu.au/HolgerBaumgardt/globular/fits/ter9.html}, although the source data and publication is not clear, presumably high-resolution spectroscopy.} $\sim7{\rm \,km\,s^{-1}}$. A comparison between both RVD agree with each other, showing a significant contamination of field stars with RV similar to the cluster, making the cluster-field decontamination a hard task, as also reported by \citet{baumgardt+19}. We speculate that the apparent large velocity dispersion observed for FSR\,1776 is due to field star contamination with similar RV. If the dispersion is reduced by a factor $3\times$ resulting in $\sim10{\rm \,km\,s^{-1}}$ it would agree better with Milky Way globular clusters. High-resolution spectroscopic analysis is required to assess a more accurate cluster membership.

Last but not least, we performed blind Gaussian mixture model (GMM) fits to simulated RVDs$_{(bulge+disc+cluster)}$ containing the three components with the parameters adopted above. The details are presented in Appendix \ref{app:gmm}. We came to the conclusion that the blind GMM fits to RVDs all led to wrong solutions, even if we force to find three components. Therefore, our procedure to rely on independent external information on the bulge and disc populations has proven essential to find a residual GC-like structure in the MUSE RVD.

%
\section{Photometry, astrometry, spectroscopy}

Now that the existence of FSR\,1776 as a star cluster is confirmed by our RV analysis, we can proceed to derive its parameters from the CMDs using multi-band photometry from 2MASS, VVV, DECaPS, and Gaia.

%

\subsection{Public photometric databases}

We explored the data available for this cluster in different public photometric databases that complement each other. For example, the 2MASS catalogue \citep{skrut+06} can be better used to define the cluster centre and the upper red giant branch (RGB; including the RGB tip and slope), because those stars in the VVV database appear to be saturated ($K_s < 11$~mag). On the other hand, the VVV survey is more suitable to measure the RC magnitude, the lower RGB and the cluster reddening. PSF-fitting techniques yield better results in our highly crowded region, so we used the VVV photometry database by \citet{alonsogarcia+18}, which was extracted using such techniques and it is equivalent to the VVV catalogue by \citet{surot+19phot}. 

In the optical wavelengths, we used the shallow photometry from Gaia EDR3 and the deep photometry from the DECam Plane Survey (DECaPS, DR1) obtained through NOIR Astro Data Lab \citep{schlafly+18}. The DECaPS photometry reaches beyond the main-sequence turn-off at the distance of the bulge, using the $grizY$ bands. The DECaPS Survey covers the highly reddened regions of the southern Galactic plane with Galactic latitude $|b| < 4$ deg and longitude $-120 < l < 5$ deg.

We have generated seven CMDs with different combinations of filters for the 450 MUSE stars, when available, and have used all CMDs together to perform the isochrone fittings. We adopted the PARSEC\footnote{\url{http://stev.oapd.inaf.it/cgi-bin/cmd}} isochrones \citep{bressan+12} in the following analysis.

%

\subsection{Extinction}

\cite{clarkson+08} has fitted multiple isochrones to the PM cleaned bulge CMD. We adopted three of those isochrones that are shaped like an envelope around the total bulge CMD as a reference to fit the specific extinction for the FSR\,1776 region. The isochrones by \cite{clarkson+08} were fixed at the bulge distance of 7.24\,kpc, whereas age and metallicity combinations were (14Gyr,-1.0dex), (11Gyr,+0.0dex), (14Gyr,+0.4dex). We adopted the extinction curve of \cite{cardelli+89} with a correction obtained by \cite{odonnell94} with $R_v = 3.1$ and coefficients given by PARSEC models for a G2V star. 

The bulge CMDs were generated using membership probability assigned to each of the 450 stars by counting how many times they are recovered when bootstrap samples are extracted from the full 450 stars sample using the mean RV, dispersion and fraction of bulge stars in the field, as defined in the previous section.

\begin{figure*}[!htb]
\centering
\includegraphics[width=0.8\textwidth]{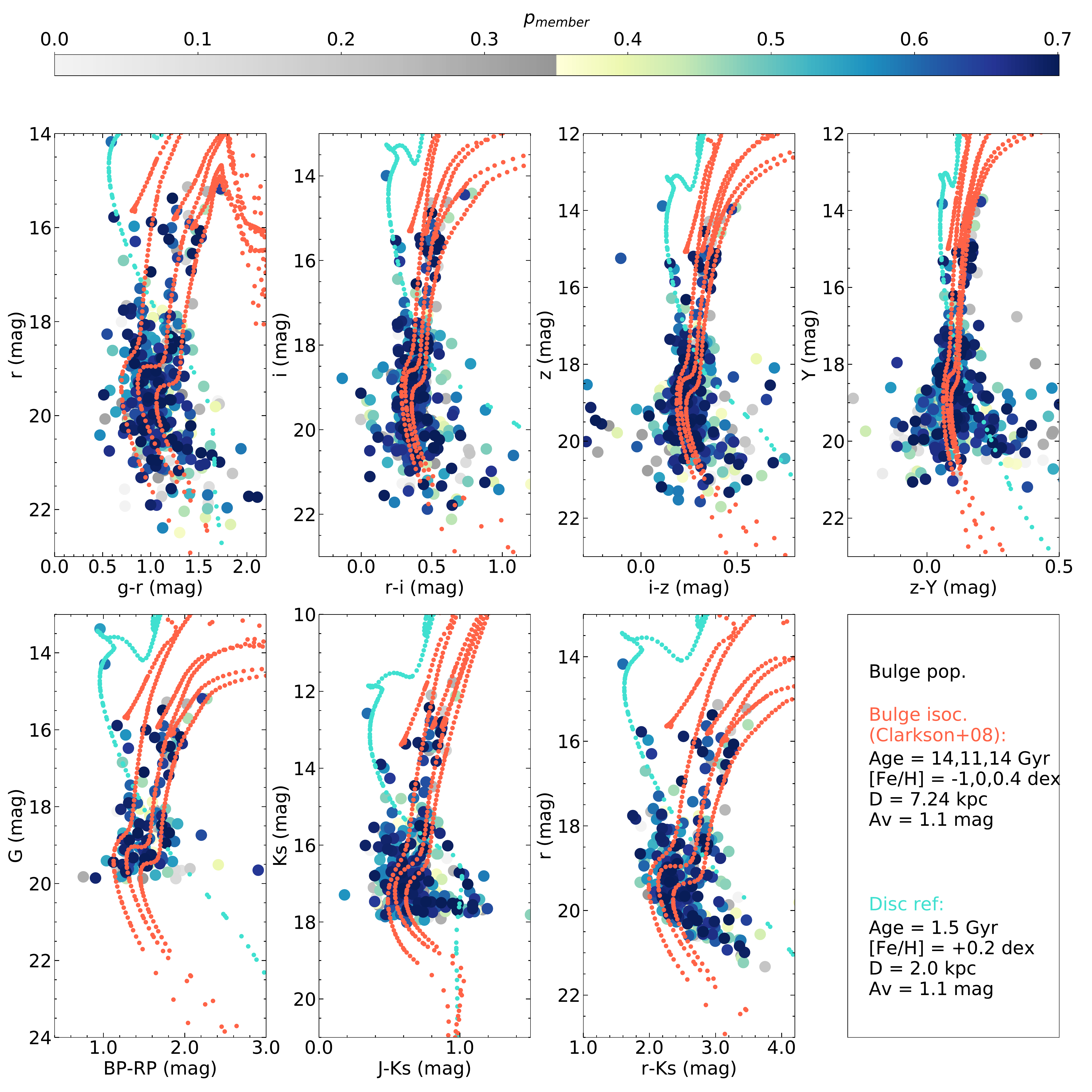}
\caption{Bulge CMD used to fit the extinction. The central isochrone with  (age,[Fe/H])=(11Gyr,0.0) should match the mean bulge CMD. The redder right isochrone with (age,[Fe/H])=(14Gyr,+0.4) should be the limit of the reddest RGB stars and the left isochrone with (age,[Fe/H])=(14Gyr,-1.0) should be the blue limit for main sequence and RGB. The cyan isochrone indicates a shorter distance and younger age to match a probable contamination from foreground disc stars in the bright blue region of all CMDs.}
\label{fig:CMDbulge}
\end{figure*}

We performed a visual isochrone fitting to all seven bulge CMDs simultaneously, varying only the extinction A$_{\rm V}$. The best fitting was obtained for A$_{\rm V} = 1.1$~mag, as shown in Fig. \ref{fig:CMDbulge}. This is consistent with the existing reddening maps, for example, the BEAM calculator \citep{gonzalez+12} yields A$_{\rm V} = 1.18$~mag for the same coordinates. Furthermore, the higher spatial resolution extinction map by \citet{surot+20ext} reveals that the foreground field extinction is fairly uniform, with a small variation in reddening, limited to $\Delta E(J-Ks) < 0.10$ mag within 10 arcmin about the cluster centre.

%
\subsection{Spectroscopic metallicities}

\begin{figure}[!htb]
    \centering
    \includegraphics[width=\columnwidth]{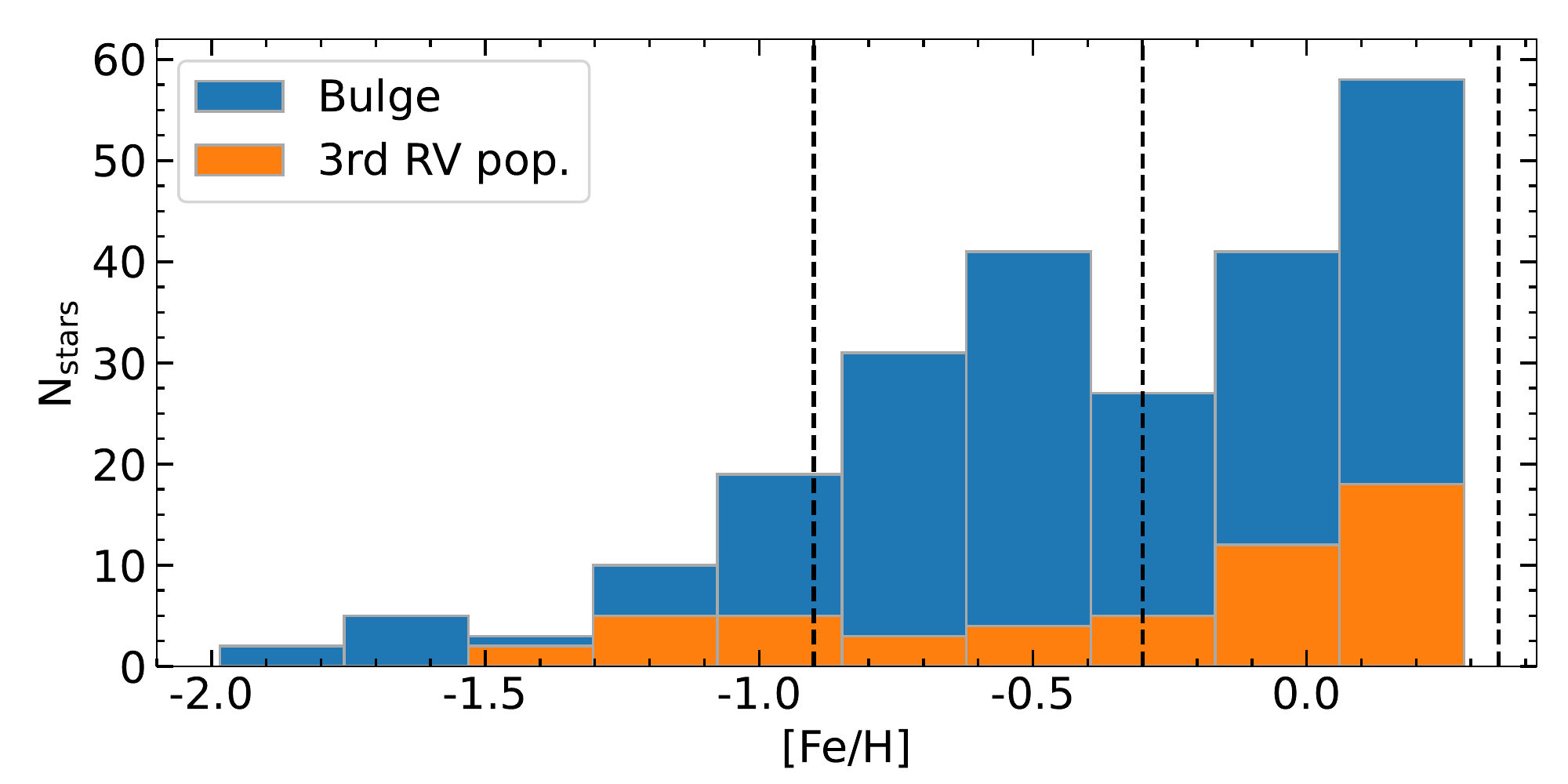}
    \caption{Metallicity distribution of a random extracted sample from the RV distribution from the bulge and cluster populations (see previous section for details. Vertical lines are the limits that define metal-rich and metal-poor peaks, visually estimated.}
    \label{fig:histmet}
\end{figure}

As explained above, the membership probabilities were estimated for all stars and for each of the three RV components based on the RV information only. However, a globular cluster should present not only a peak in the RVD but also a peak in the [Fe/H] distribution. Therefore, as a second step, we look for a metallicity peak within the RV-selected cluster stars. We show the [Fe/H] distribution for the third RV component in Fig. \ref{fig:histmet} with a clear peak at solar metallicity and above. The third RV peak detected as the signature of FSR\,1776 presents some contamination from bulge stars because both RV distributions overlap. In particular, when we extract a sub-sample of bulge and cluster stars following the procedure explained above, both metallicity distributions show two peaks (see Fig. \ref{fig:histmet}). In order to check whether the prominent metal-rich peak of the third RV component is really the cluster, we perform some tests.  We have extracted 1000 bootstrapped samples and compared bulge and cluster metallicity distributions. Bulge stars have about $\sim$15\% more metal-rich stars compared to the metal-poor stars, whereas for the cluster, the ratio is of about $\sim$30\% more metal-rich stars. Therefore, the third RV peak a.k.a. FSR\,1776, reveals an excess of metal-rich stars of about $\sim$15\% compared to what is expected for the bulge population. We interpret this as an indication that FSR\,1776 has a single peak metallicity around solar metallicity. This result must be confirmed by a high-resolution spectroscopic analysis over a larger area around the cluster position.

%

\subsection{FSR\,1776 CMD isochrone fitting}

\begin{figure*}[!htb]
\centering
\includegraphics[width=0.8\textwidth]{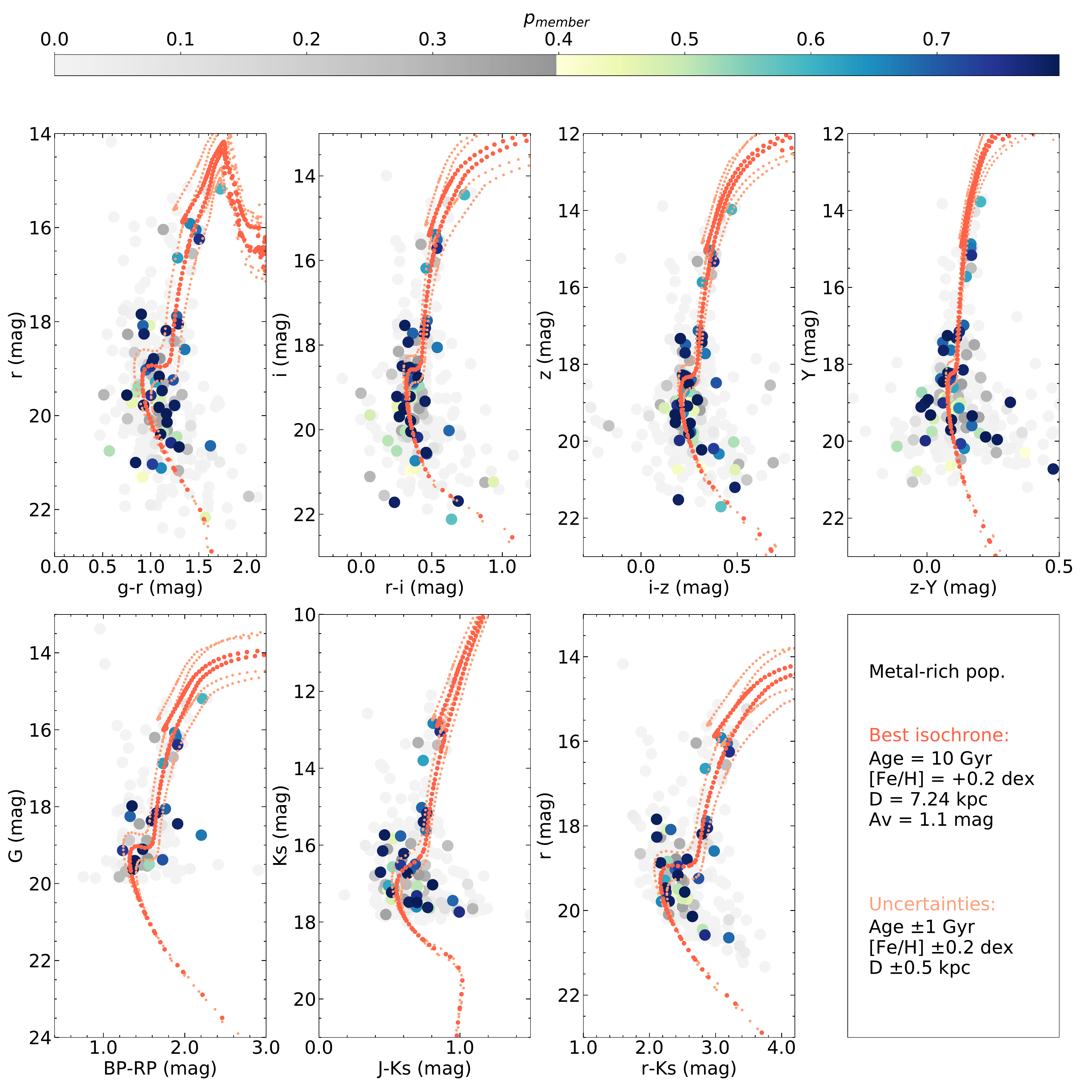}
\caption{Cluster CMD (MR component). The red isochrone is the best fit to the data, and the isochrones bracketing the estimated uncertainties are also displayed.}
\label{fig:CMDclusterMR}
\end{figure*}

\begin{figure*}[!htb]
\centering
\includegraphics[width=0.8\textwidth]{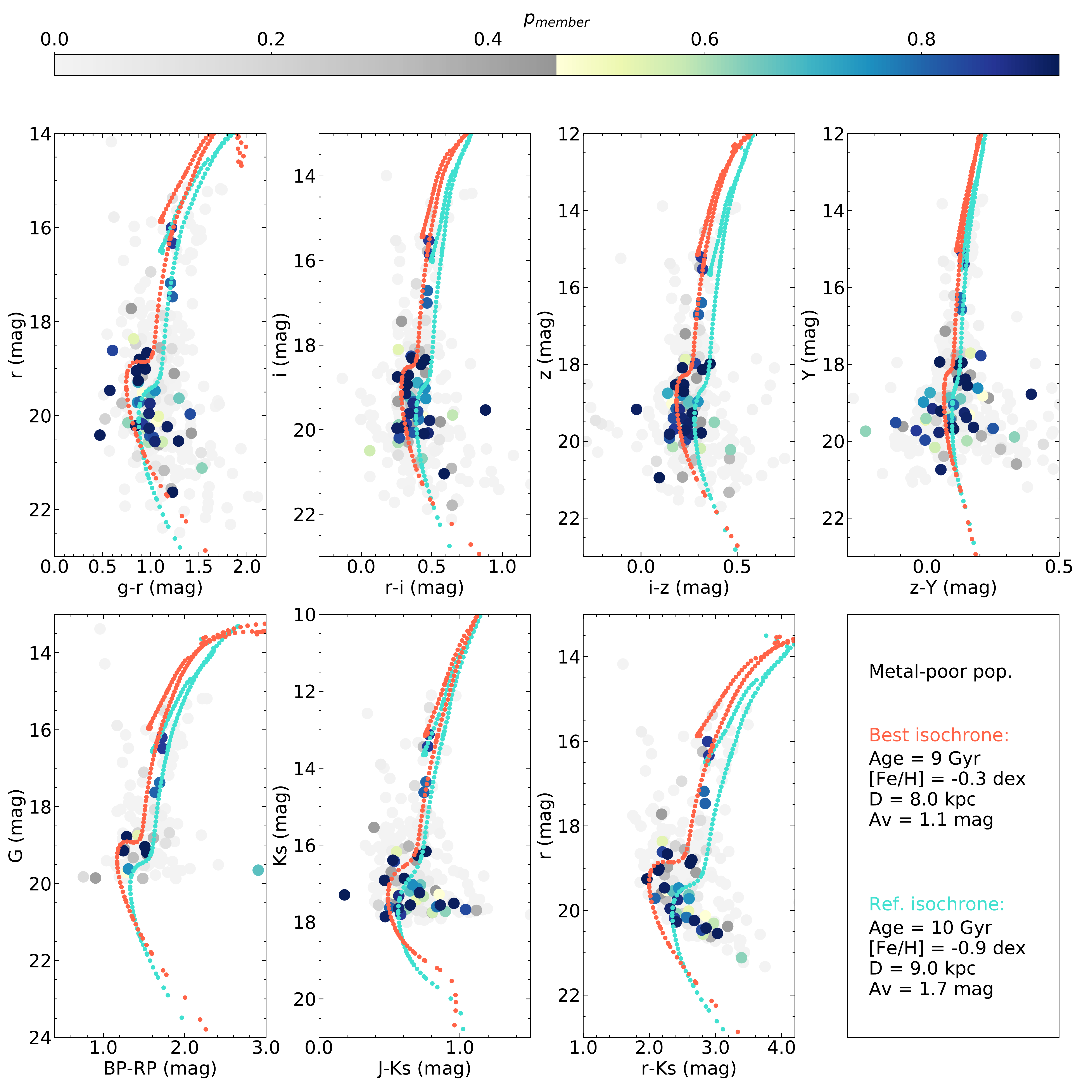}
\caption{Cluster CMD (MP component). The cyan reference isochrone is the best fit obtained in the NIR CMD by forcing a solution for very low metallicity. While this is possible by increasing the extinction, the fits fail in the other CMDs. The red isochrone is the best fit to the data.}
\label{fig:CMDclusterMP}
\end{figure*}

The same procedure used to find a RV-cleaned bulge CMD is applied to find the cluster CMD. As discussed above, the RV-cleaned cluster sample has some bulge contamination and it seems that the cluster is represented by the metal-rich peak. Nevertheless, in order to keep the CMD isochrone fitting as unbiased as possible, we assign cluster membership probabilities for all stars based only on RV and split them into metal-rich (MR, $-0.3 < {\rm [Fe/H]} < +0.5$) and metal-poor (MP, $-0.9 < {\rm [Fe/H]} < -0.3$) components to proceed with parallel analysis (see Fig. \ref{fig:histmet}). For details on the spectroscopic metallicity derivation, see Appendix \ref{app:etoile}. The isochrone fitting results for both cases seem reasonable, as it can be seen in Figs. \ref{fig:CMDclusterMR} and \ref{fig:CMDclusterMP}. A possible way to decide which CMD corresponds to the cluster is to analyse their PMs.

%

\subsection{Proper motions}

Gaia EDR3 does not provide a very deep photometry as it can be appreciated in Figs. \ref{fig:CMDbulge}, \ref{fig:CMDclusterMR}, \ref{fig:CMDclusterMP}, which means that only a handful of RGB stars can be used. Even though this low-number statistics cannot provide a final word on the FSR\,1776 PMs, it is possible to find a first approximation for its angular velocity on sky.

\begin{figure}[!htb]
\centering
\includegraphics[width=0.9\columnwidth]{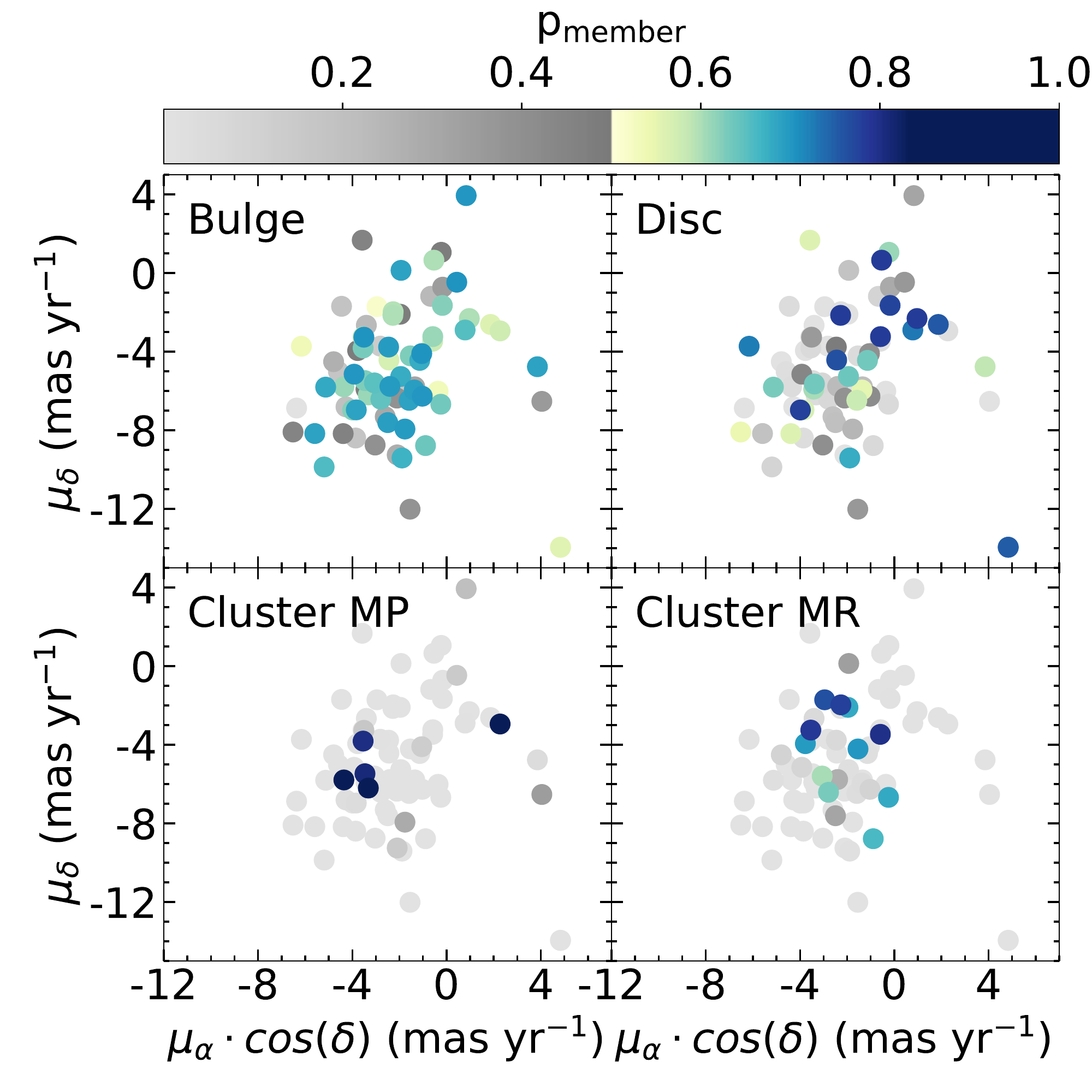}
\caption{Proper motions for all stars in common between MUSE and Gaia EDR3 with errors in PMs better than $0.3~{\rm mas\, yr^{-1}}$. The same points are shown in the four panels, but the colours of the points refer to the probability of a star's belonging to each population, as calculated from the MUSE RVD.}
\label{fig:PMbulgedisc}
\end{figure}

We show the PM distributions for bulge, disc, cluster MP and cluster MR components in separate panels in Fig. \ref{fig:PMbulgedisc}. It is possible to visually identify the loci of the bulge, disc, cluster MP and cluster MR components where the stars with the highest membership probabilities are concentrated in the PMs space. The components are located roughly at ($\langle\mu_{\alpha}\rangle,\langle\mu_{\delta}\rangle$) $\sim$ (-3,-6), (0,-2), (-4,-6), (-2,-2) ${\rm mas\, yr^{-1}}$, respectively. Bulge and disc are clearly separated. The cluster MP component seems to match the locus of the bulge, whereas the cluster MR component seems to be in a different locus from those loci of the bulge and disc. Therefore, we conclude that FSR\,1776 is the cluster MR component. Calculating the average of the PMs only of the stars with a membership probability higher than 70\%, we find ($\langle\mu_{\alpha}\rangle,\langle\mu_{\delta}\rangle$) $=$ ($-2.3\pm1.1,-2.6\pm0.8$) ${\rm mas\, yr^{-1}}$ for FSR\,1776.

%

\section{Cluster orbit}

The orbits have been computed with the GravPot16\footnote{https://gravpot.utinam.cnrs.fr} model by adopting the same Galactic configuration as described in \cite{fertrinc+20}, except for the bar, which we adopted 41 km\,s$^{-1}$\,kpc$^{-1}$ as the bar pattern speed as suggested by \cite{sanders+19b}.

\begin{figure}[!htb]
\centering
\includegraphics[width=0.9\columnwidth]{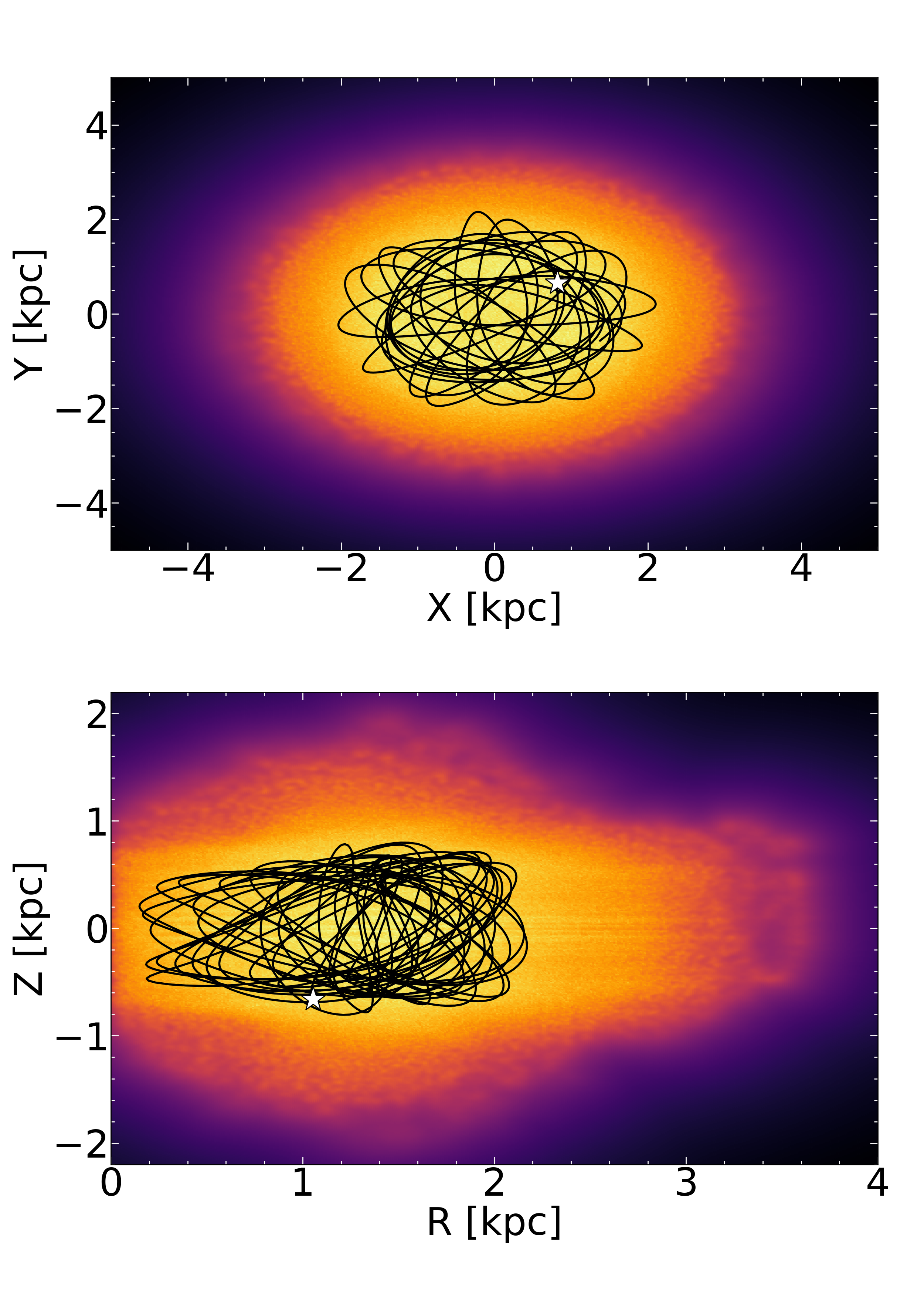}
\caption{Probability density in the equatorial Galactic plane (top)
and side-on (bottom) of 100,000 simulated orbits of FSR\,1776 time-integrated backward for 1 Gyr. Yellow (orange) colours correspond to more probable regions of the space more frequently crossed by the simulated orbits. The black line refers to the cluster orbit computed with the central observable. The white ‘star’ symbol marks the present-position of FSR\,1776.}
\label{fig:orbits}
\end{figure}

Figure \ref{fig:orbits} shows the X-Y projection (face-on orbit) and Z-R projection (meridional orbit) of the FSR\,1776 orbits. The black line in the same figure shows the central orbit of the cluster by adopting the central input parameters (e.g., RA, DEC, longitude, latitude, distance, RV, $\mu_{\alpha}$, $\mu_{\delta}$), whilst the colour maps correspond to 100,000 orbits generated by adopting a simple Monte Carlo approach that considers the errors in the input parameters as 1$\sigma$ variation.

FSR\,1776 has prograde orbits with ellipticities of 0.88$\pm$0.15, a maximum vertical height, $Z_{max}$, of $\sim$0.87$\pm$0.19 kpc, an orbital pericentre ($r_{min}$) and apocentre ($r_{max}$) radii of $\sim$0.14$\pm$0.25 kpc and $\sim$2.26$\pm$0.44 kpc, respectively, placing FSR\,1776 well within the inner bulge of the Milky Way. In addition, using the slightly different angular velocity for the bar (31, 41, and 51 km\,s$^{-1}$\,kpc$^{-1}$) does not significantly change the results and returns orbits in which the cluster is confined to the bulge region. The orbital integration, therefore, indicates that this GC belongs to the inner bulge.

\begin{figure*}[!htb]
\centering
\includegraphics[width=\textwidth]{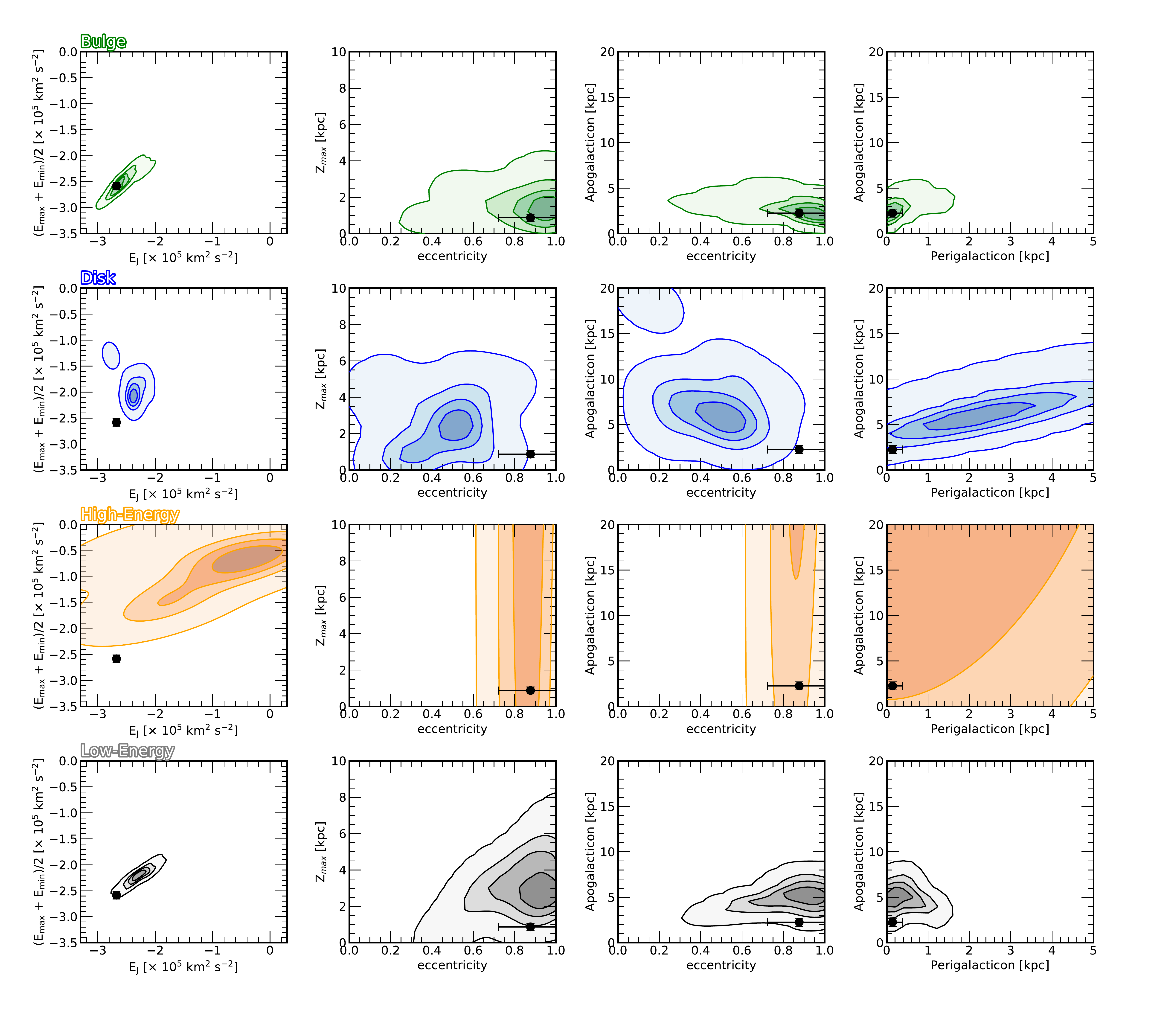}
\caption{Kernel Density Estimation (KDE) models of the characteristic orbital energy ((E$_{\rm max}$ + E$_{\rm min}$)/2), the orbital Jacobi energy (E$_{\rm J}$), orbital pericentre and apocentre, orbital eccentricity, maximum vertical height above the Galactic plane for GCs with an Galactic origin \citep[e.g.][]{massari+19}. FSR~1776 is highlighted with a black dot symbol.}
\label{fig:energybulge}
\end{figure*}

\begin{figure*}[!htb]
\centering
\includegraphics[width=\textwidth]{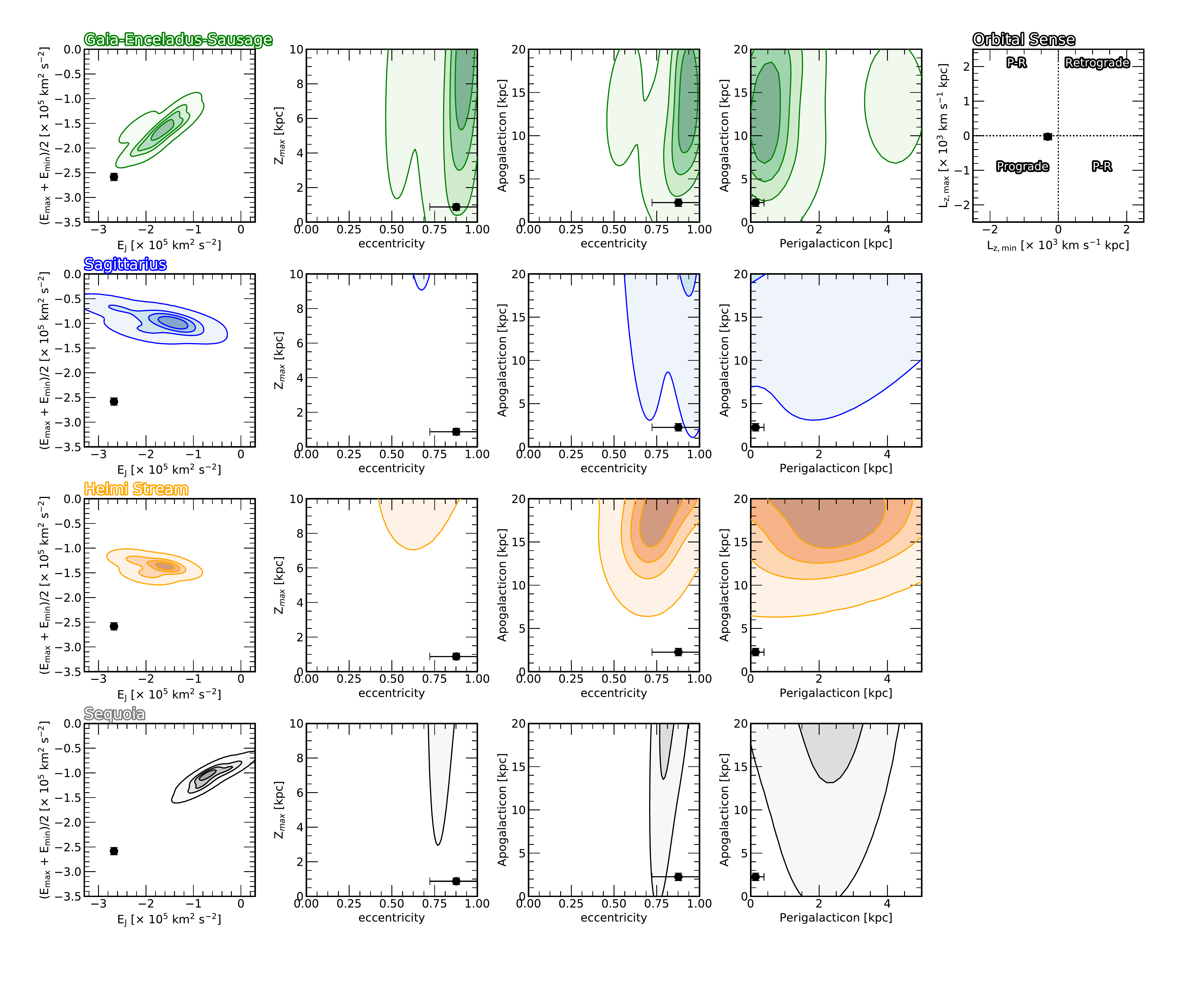}
\caption{Kernel Density Estimation (KDE) models of the characteristic orbital energy ((E$_{\rm max}$ + E$_{\rm min}$)/2), the orbital Jacobi energy (E$_{\rm J}$), orbital pericentre and apocentre, orbital eccentricity, maximum vertical height above the Galactic plane for GCs with an accreted origin \citep[e.g.][]{massari+19}. FSR~1776 is highlighted with a black dot symbol. The top-right panel show the minimal and maximum value of the z-component of the angular momentum in the inertial frame, and indicates the regions dominated by prograde and retrograde orbits, and those dominated by orbits that change their sense of motion from prograde to retrograde (P-R). }
\label{fig:energystreams}
\end{figure*}

Figure \ref{fig:energybulge} reveals that FSR~1776 exhibits orbital elements with high probability to belong to the family of GCs with a Galactic origin, in particular associated with the group of Main Bulge GCs, confirming that FSR~1776 is a bulge GC. Figure \ref{fig:energystreams} shows that the orbital elements of FSR~1776 have low probability to belong to the family of GCs with an accreted origin.

%

\section{Discussion}

While there is good agreement in the distance determinations for this cluster, the same does not hold for age estimations. Indeed, age values derived for FSR\,1776 are controversial, since they range from $3.2$ Gyr \citep{kharch+16} to larger than $10$ Gyr, as suggested by \citet{minniti+17a} and \citet{palma+19}, mainly because of the presence of RR\,Lyrae stars in the cluster projected field. The RR\,Lyrae variable stars ``are unequivocally old ($\geq10$ Gyr)'' \citep{kunder+18,beaton+18}, being excellent tracers of an old stellar population. Analysing the distances of the five RR\,Lyrae found in the field of this cluster \citep[7.8, 8.3, 8.9, 9.5 and 9.7 kpc,][]{palma+19}, we cannot classify any of them as a cluster member, because they are all farther than 0.5\,kpc from the estimated distance of FSR\,1776 (7.24$\pm$0.5\,kpc, see Fig. \ref{fig:CMDclusterMR}), even though they are all within $3.5\arcmin$ from the cluster projected centre; therefore, they should be background RR\,Lyrae belonging to the bulge. In fact, the metallicity of FSR\,1776 is [Fe/H]$_{phot} = + 0.2\pm$0.2 or [Fe/H]$_{spec}=~+0.02\pm0.01~(\sigma~=~0.14$~dex). For such high metallicities, only lighter stars from the main sequence (MS) would become RR\,Lyrae \citep[e.g.][]{marconi+15}, i.e., it would take more than a Hubble time for these stars to pulsate as RR~Lyrae, and FSR\,1776 is only 10$\pm$1\,Gyr old. Therefore, the fact that no RR Lyrae members are found in this cluster is consistent with the estimated age and metallicity for this GC.

\begin{figure}[htb]
\includegraphics[width=0.9\columnwidth]{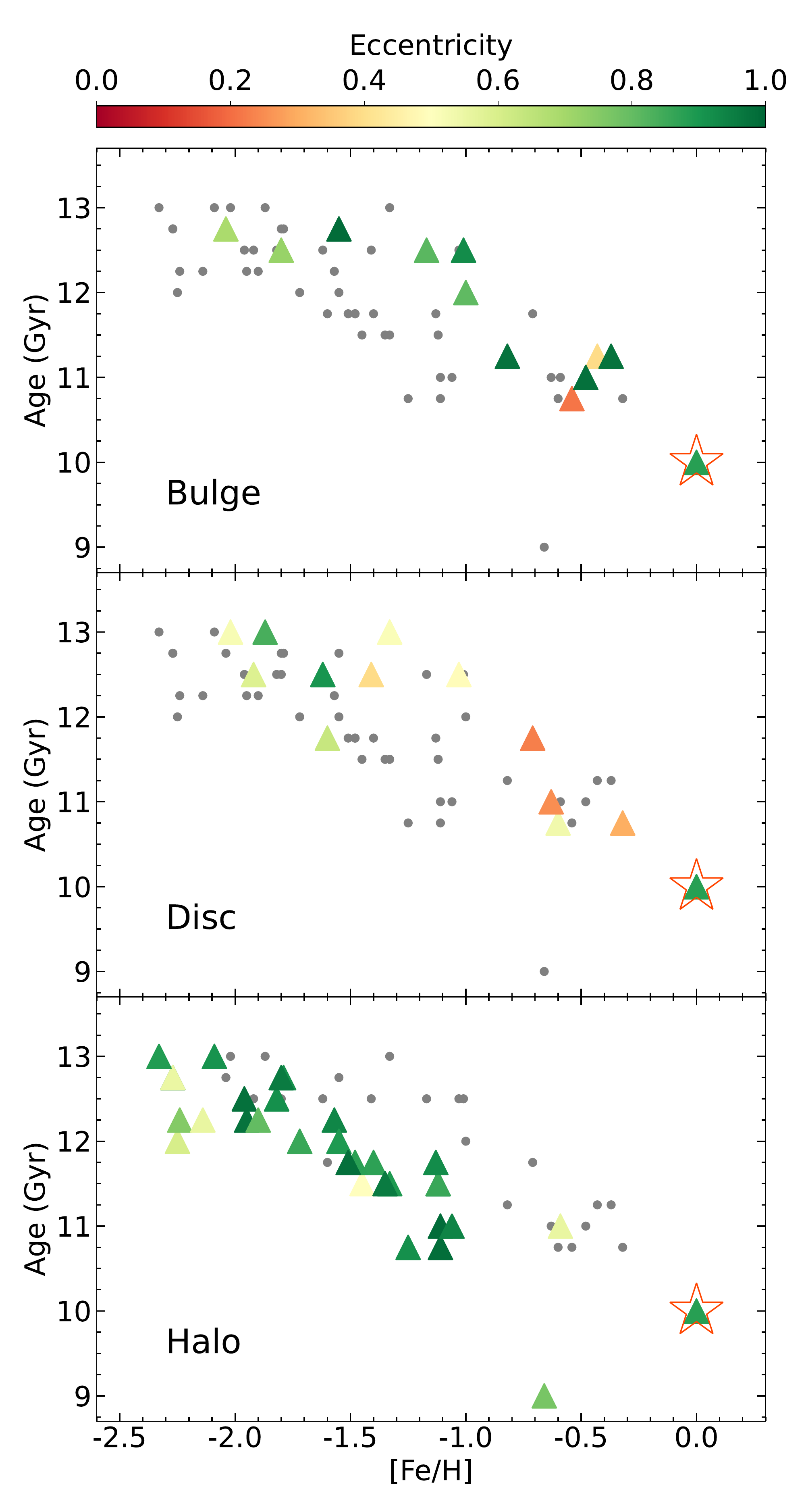}
\caption{Milky Way GCs age-metallicity relation using homogeneous ages from \citet{vandenberg+13} and average metallicities 
(\protect\url{ http://www.sc.eso.org/~bdias/catalogues.html}) 
on the metallicity scale by \citet[][2019 edition]{dias+16}. Eccentricity comes from the calculations from \texttt{GravPot16}. Bulge, disc, and halo classification is adopted from \citet{dias+16}. FSR\,1776 is identified with a star contour.}
\label{fig:amr}
\end{figure}

\begin{figure}
    \centering
    \includegraphics[width=\columnwidth]{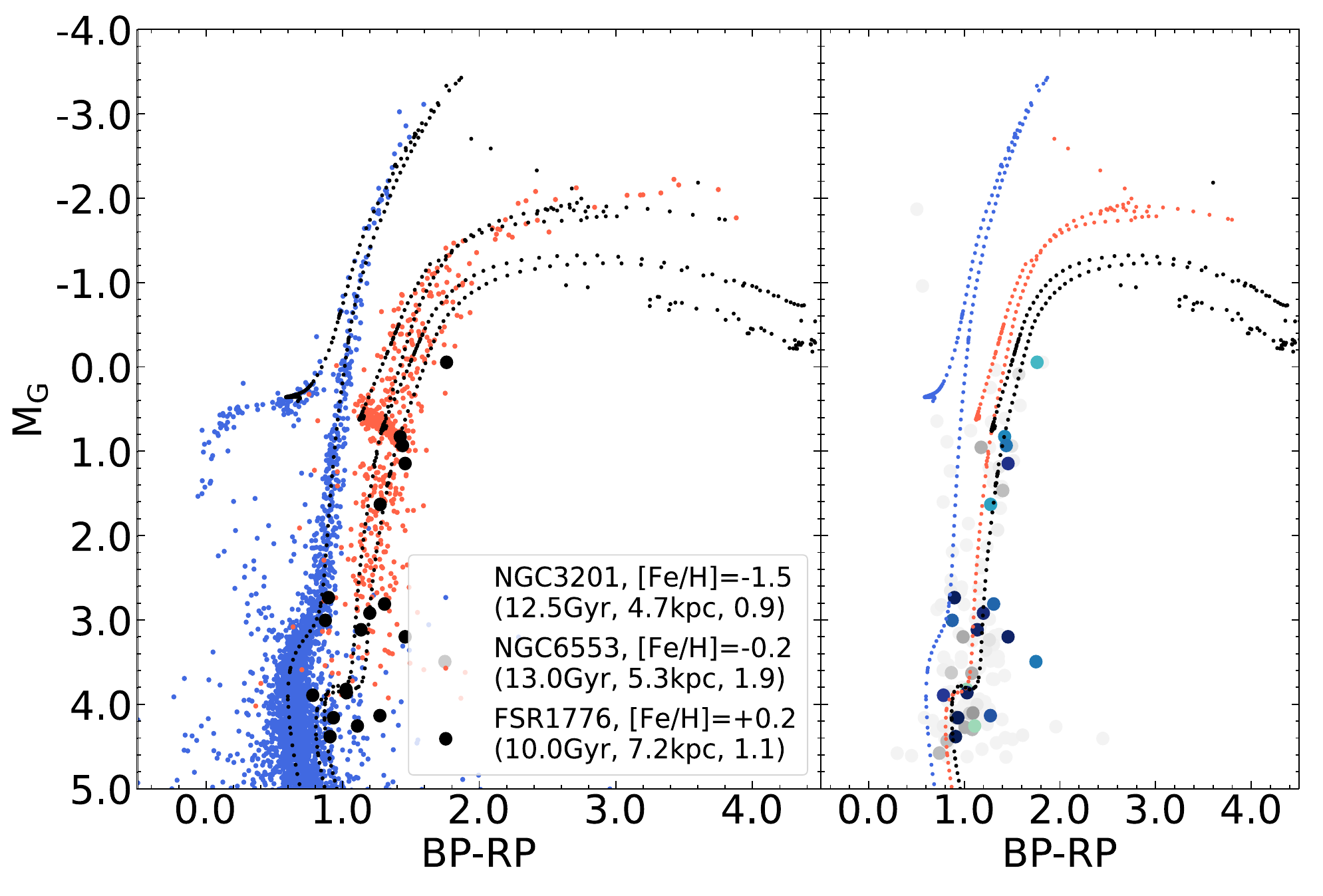}
    \caption{Gaia EDR3 CMD, proper-motion cleaned, of the well-known old GCs NGC\,3201 and NGC\,6553 as proxies for the locus of metal-poor and metal-rich globular clusters. Black points on the left panel are FSR\,1776 stars with $p_{mem} > 40\%$ whereas the right panel shows all stars colour-coded as in Fig. \ref{fig:CMDclusterMR} to ease the visualisation. PARSEC isochrones are the same on the left and right panels with parameters from the literature.}
    \label{fig:gaiacomp}
\end{figure}

\cite{vandenberg+13} have derived homogeneous ages for 55 Milky Way GCs. \cite{leaman+13} showed that they follow two different paths in the age-metallicity relation (AMR) diagram, the more metal-poor path corresponding to halo clusters, and the more metal-rich path corresponding to disc and bulge clusters overlapped. \cite{massari+19} have discussed the Milky Way GC AMR also using orbital parameters showing that the metal-rich path has indeed a mix of populations. We show in Fig. \ref{fig:amr} the AMR for Milky Way GCs where different populations are identified in different panels for clarity, with colours representing the orbital eccentricity that seems to split well the general behaviour of bulge and disc clusters, with the former having mostly eccentric orbits and the latter having mostly circular orbits (see also Fig. \ref{fig:energybulge}). FSR\,1776 is added to these panels with the age, metallicity, and eccentricity derived in the present work. It can be observed in Fig. \ref{fig:amr} that FSR\,1776 follows well the extrapolation of the bulge AMR, even though the bulge has a complex formation history and the AMR is more disperse (see e.g. \citealp{moni-bidin+21}). FSR\,1776 is among the youngest and most metal-rich bulge GCs known so far (see also Fig. \ref{fig:gaiacomp} for a direct comparison of CMDs with known old GCs with different metallicities).

%

\section{Summary and conclusions}

We have used MUSE data to verify the GC nature of FSR\,1776. The RVD revealed a residual population when bulge and disc are subtracted, which is the confirmation of FSR\,1776 as a cluster with RV $= -103.7\pm 0.4~{\rm km}\,{\rm s}^{-1}$. In a second step, multi-band photometry and astrometry have been used to fully derive its parameters. A sample of stars with the cluster RV is still contaminated by bulge stars with different metallicities. A statistical comparison with the bulge metallicity distribution in addition to the PM distribution were used to conclude that the metal-rich component is the unique population different from bulge and disc, with an average estimated as [Fe/H]$_{spec}=~+0.02\pm0.01~(\sigma~=~0.14$~dex) and  ($\langle\mu_{\alpha}\rangle,\langle\mu_{\delta}\rangle$) $=$ ($-2.3\pm1.1,-2.6\pm0.8$) ${\rm mas\, yr^{-1}}$. The isochrone fitting provided for FSR~1776 a distance of 7.24$\pm$0.5~kpc, an age of 10$\pm$1~Gyr, a metallicity $[{\rm Fe/H}]=+0.2\pm$0.2, and an extinction $A_V\approx1.1$~mag.

The orbits revealed that FSR\,1776 is confined within the inner Galactic bulge, at least for the past 1~Gyr. Its age and metallicity are consistent with the bulge AMR extrapolation. FSR\,1776 is populating the AMR locus around $\sim$10\,Gyr and solar metallicity together with new discoveries such as Patchick\,99 \citep{garro+21}, FSR\,19 and FSR\,25 \citep{obasi+21}, and they may be the missing link between typical GCs and the metal-rich bulge field.

Deeper spatially resolved photometry and high-resolution spectroscopy of a larger number of stars is required to separate cluster and field stars with higher accuracy and fully characterise FSR\,1776. The photometry will be useful for statistical CMD decontamination as well as PM for fainter stars. The spectroscopy will be important to reduce uncertainties in RV, metallicities and get the first chemical abundance determinations for FSR\,1776, for example [$\alpha$/Fe].

%

\begin{acknowledgements}
B.D. is grateful to S. Kamann for useful discussions and support throughout the MUSE data analysis using PampelMUSE.\\
Based on observations collected at the European Southern Observatory under ESO programme 0101.D-0363(A).\\
We gratefully acknowledge data from the ESO Public Survey program ID 179.B-2002 taken with the VISTA telescope, and products from the Cambridge Astronomical Survey Unit (CASU).\\
This publication makes use of data products from the Two Micron All Sky Survey, which is a joint project of the University of Massachusetts and the Infrared Processing and Analysis Center/California Institute of Technology, funded by the National Aeronautics and Space Administration and the National Science Foundation.\\
This work has made use of data from the European Space Agency (ESA) mission
{\it Gaia} (\url{https://www.cosmos.esa.int/gaia}), processed by the {\it Gaia}
Data Processing and Analysis Consortium (DPAC,
\url{https://www.cosmos.esa.int/web/gaia/dpac/consortium}). Funding for the DPAC
has been provided by national institutions, in particular the institutions
participating in the {\it Gaia} Multilateral Agreement.\\
This research uses services or data provided by the Astro Data Lab at NSF's National Optical-Infrared Astronomy Research Laboratory. NOIRLab is operated by the Association of Universities for Research in Astronomy (AURA), Inc. under a cooperative agreement with the National Science Foundation.\\
T.P. and J.J.C. acknowledge support from the Argentinian institution SECYT (Universidad Nacional de Córdoba).\\
D.M. acknowledges support from FONDECYT Regular grants No. 1170121. 
J.A.-G. acknowledges support from Fondecyt Regular 1201490 and from ANID – Millennium Science Initiative Program – ICN12\_009 awarded to the Millennium Institute of Astrophysics MAS.\\
B.B. acknowledges partial financial support from FAPESP, CNPq and CAPES - Finance code 001.\\
R.K.S. acknowledges support from CNPq/Brazil through project 305902/2019-9.
\end{acknowledgements}

%

  \bibliographystyle{aa.bst} 
  \bibliography{minni23.bib} 

\appendix

%
\section{Validation of the full spectrum fitting method for all spectral types}
\label{app:etoile}

\begin{figure}[!htb]
\centering
\includegraphics[width=\hsize]{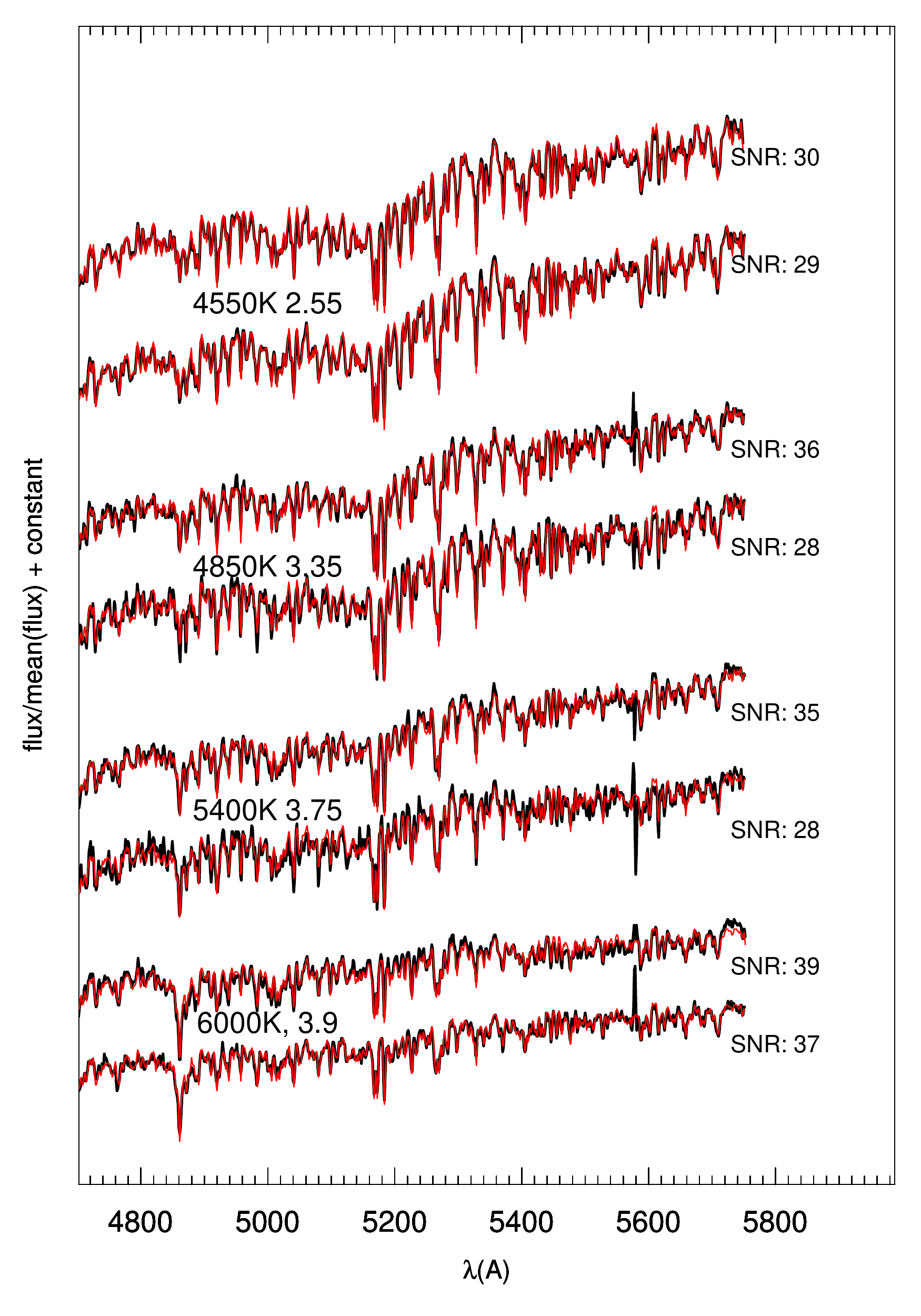}
\caption{Example of spectra in the visible region, used to fit the atmospheric parameters, of eight selected stars from the FSR\,1776 cube displayed as black lines. The red line is the best fit template spectrum from the MILES library. We give examples in a sequence from top to bottom of increasing surface gravity and temperature. For each combination of these two parameters (text on the left with the approximate $T_{\rm eff}$ and log($g$)) we show a pair of spectra with higher and lower SNR and magnitude, all of them with similar [Fe/H]$\sim+0.1$. From top to bottom, the pairs represent the RC, lower RGB, SGB, and MS turnoff.}
\label{fig:spec}
\end{figure}

\cite{dias+15,dias+16} described in detail the method we used here to derive atmospheric parameters. They validated this method only for RGB stars, which worked very well and produced T$_{\rm eff}$, log($g$), and [Fe/H] in very good agreement with those from high-resolution spectroscopy. In this paper, we extended the application of the method also to sub-giant branch (SGB) stars, so we present a validation of the method also for SGB and MS stars.
An example of the visible portion of the spectra around the MgI triplet of  selected stars spanning different parameters and SNR is shown in Fig. \ref{fig:spec}. We also over plot a library reference spectrum to stress the high SNR of the extracted spectrum showing many atomic and molecular lines.

\begin{figure*}[htb]
\begin{minipage}[b]{0.52\linewidth}
\centering
\includegraphics[width=\hsize]{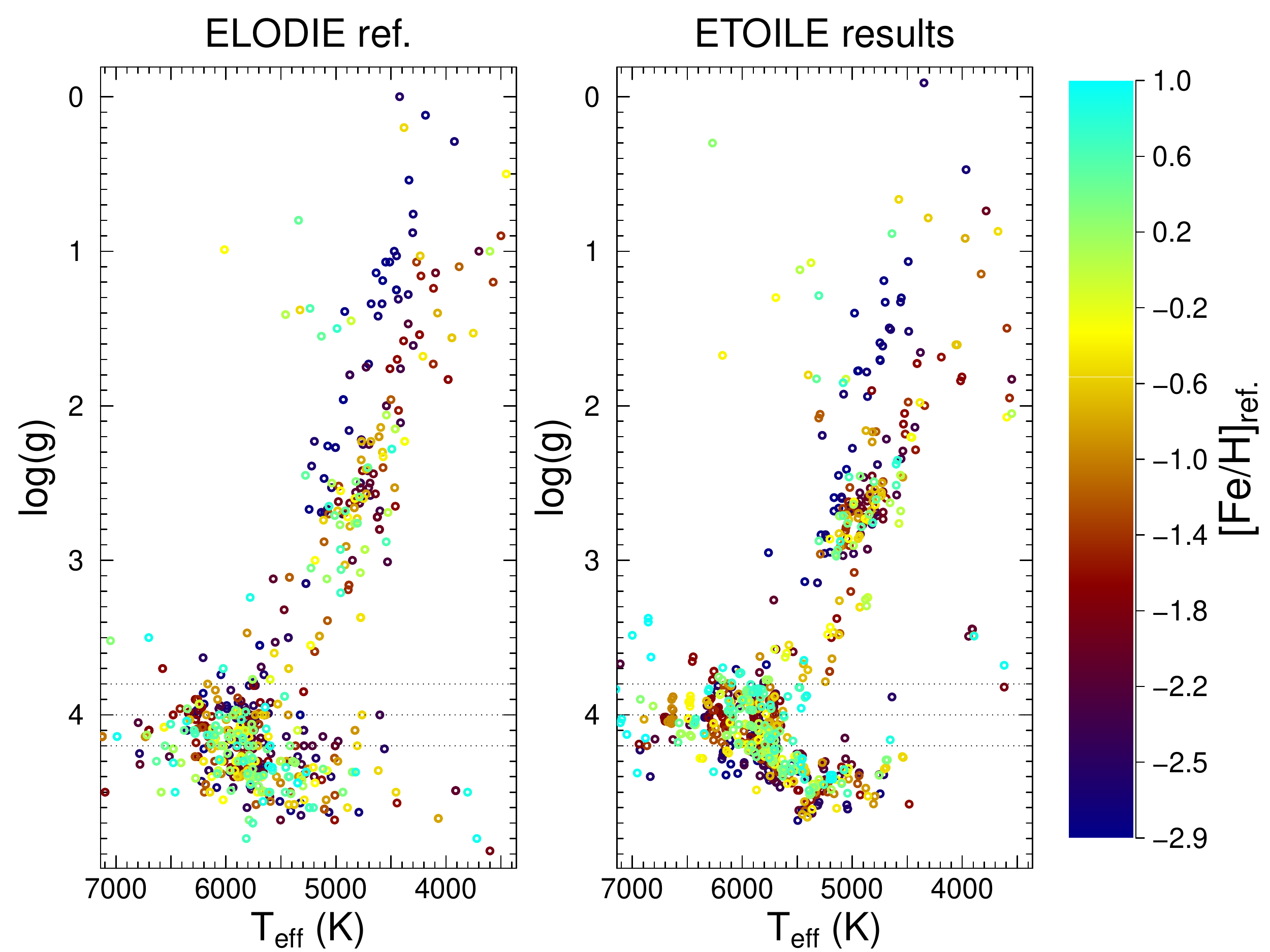}
\end{minipage}
\hspace{0.5cm}
\begin{minipage}[b]{0.38\linewidth}
\centering
\includegraphics[width=\hsize]{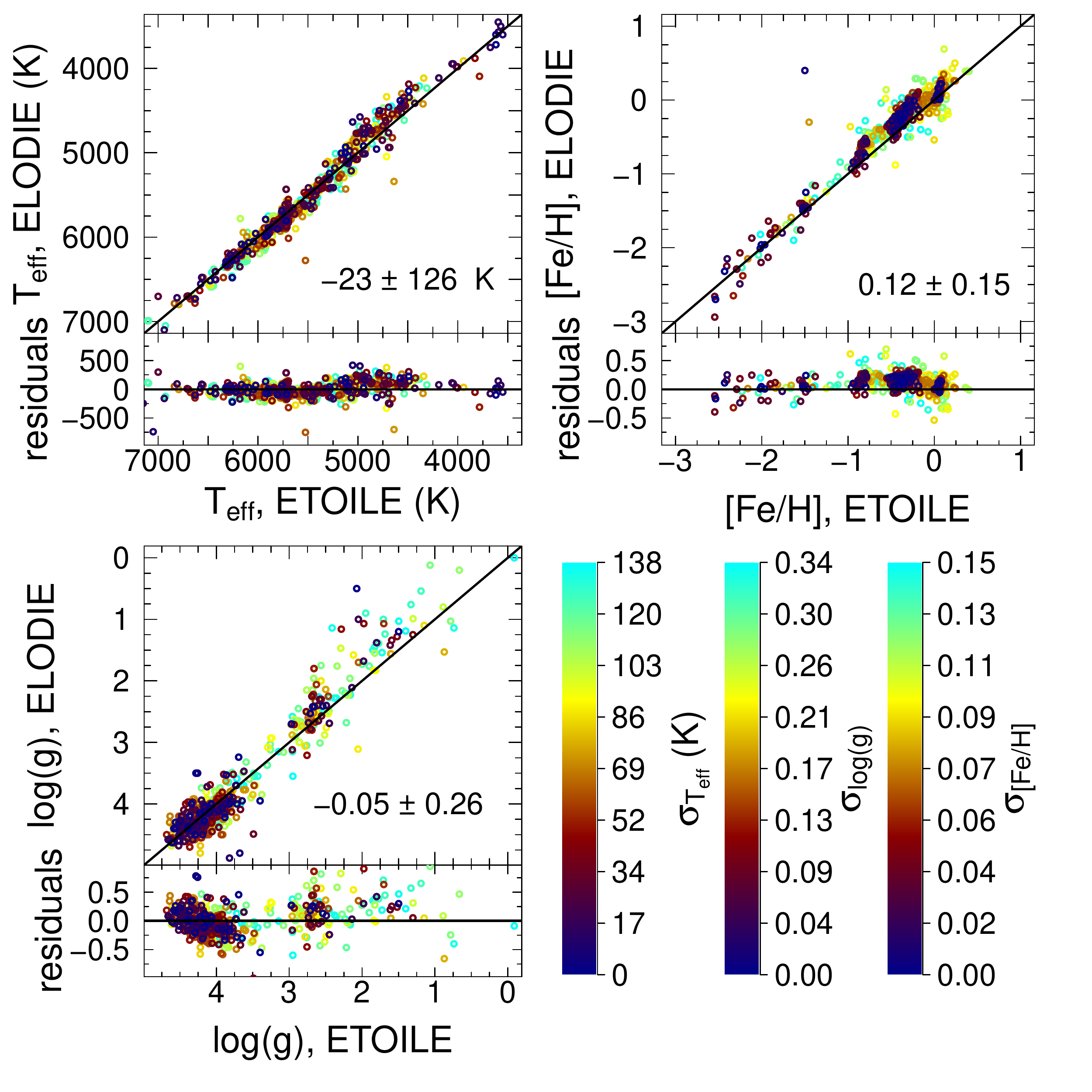}
\end{minipage}
\caption{{\it Left:} HRD with average parameters from the literature done by the ETOILE team before selecting the targets. Colours indicate the metallicity. The HRD labelled "ETOILE" is made of T$_{\rm eff}$ and log($g$) derived in this paper for the exact same stars and colours and represent the reference metallicities for a better visualisation. {\it Right:} Residuals of the one-to-one comparison between the reference and derived parameters.
Colours represent the uncertainties of the derived values for each parameter.}
\label{fig:elodie}
\end{figure*}

\begin{figure*}[htb]
\begin{minipage}[b]{0.52\linewidth}
\centering
\includegraphics[width=\hsize]{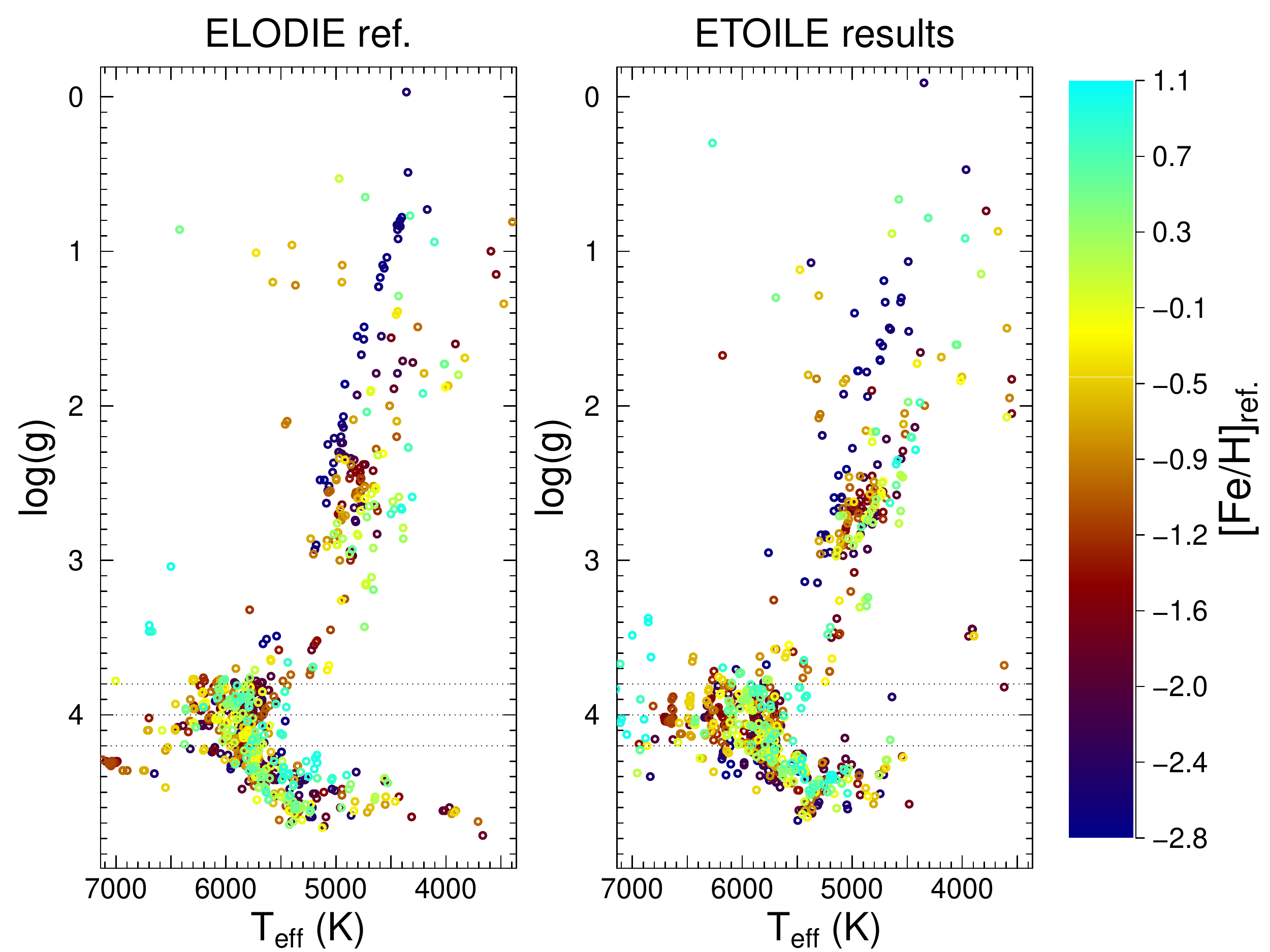}
\end{minipage}
\hspace{0.5cm}
\begin{minipage}[b]{0.38\linewidth}
\centering
\includegraphics[width=\hsize]{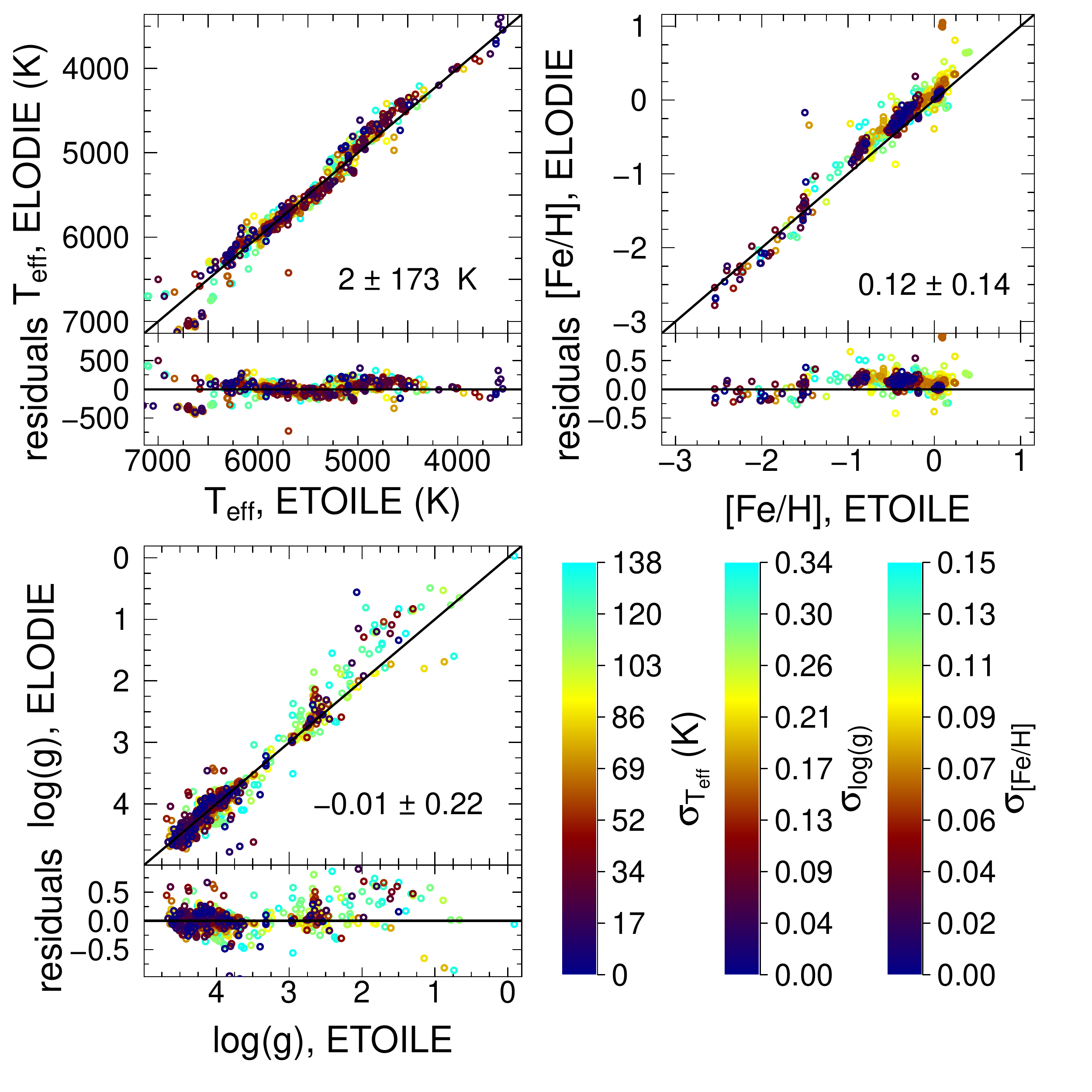}
\end{minipage}
\caption{Same as Fig. \ref{fig:elodie} but the ELODIE reference parameters are now those derived by the ELODIE team using the TGMET code \citep{katz+98} and their own library.}
\label{fig:elodieM}
\end{figure*}

The ELODIE library \citep[][2007 edition]{prugniel+01} provides 1962 medium-resolution ($\Delta\lambda = 0.55{\rm\AA}$) flux-calibrated spectra for 1388 well-known stars covering the parameter space $3100 K < \,$T$_{\rm eff} < 50000 K$, $-0.25 < $log($g$)$ < 4.9$, and $-3 <\,$ [Fe/H] $< +1$. We convolved all the spectra to the resolution of MUSE ($\Delta\lambda = 2.5{\rm\AA}$) and derived their atmospheric parameters following the recipes by \cite{dias+15}. We compared our results with those provided by the ETOILE team \citep{prugniel+07} in Figs. \ref{fig:elodie} and \ref{fig:elodieM}. The reference parameters provided by ELODIE on Fig. \ref{fig:elodie} were averaged out from the literature as described in \cite{prugniel+01}. Our results agree very well with the reference parameters within a dispersion of 126 K, 0.26 dex, and 0.15 dex for T$_{\rm eff}$, log($g$) and [Fe/H], respectively. The reference parameters presented in Fig. \ref{fig:elodieM} were derived by the ELODIE team using the TGMET code \citep{katz+98} which is an ancient version of the ETOILE code that we use here, but using their own reference spectral library. Not surprisingly the results also agree very well with similar dispersion as found in Fig. \ref{fig:elodie}, namely 173 K, 0.22 dex, and 0.14 dex for T$_{\rm eff}$, log($g$) and [Fe/H], respectively.

%
\section{Gaussian mixture model tests}
\label{app:gmm}

\begin{figure}[htb]
\centering
\includegraphics[width=0.8\columnwidth]{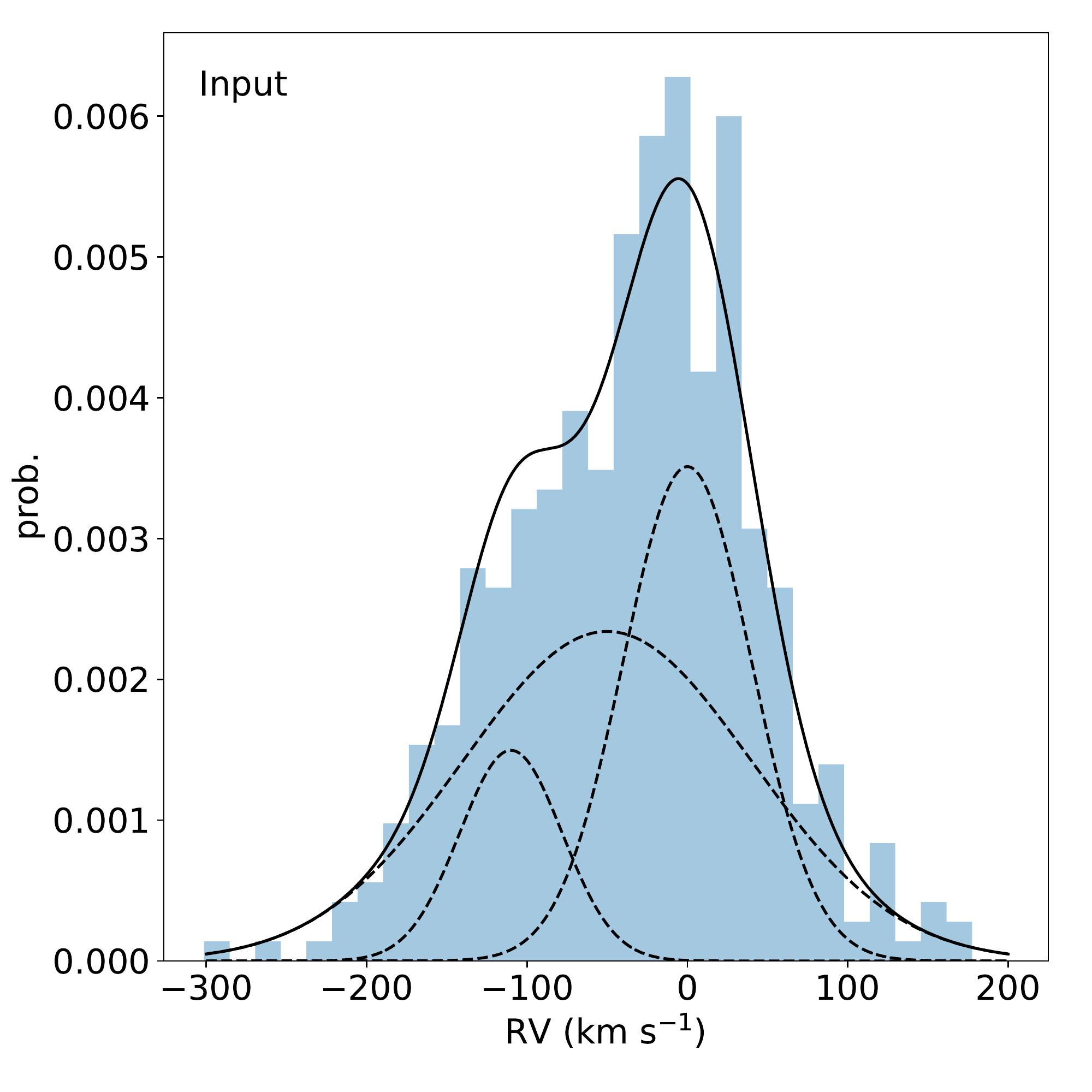}
\includegraphics[width=0.8\columnwidth]{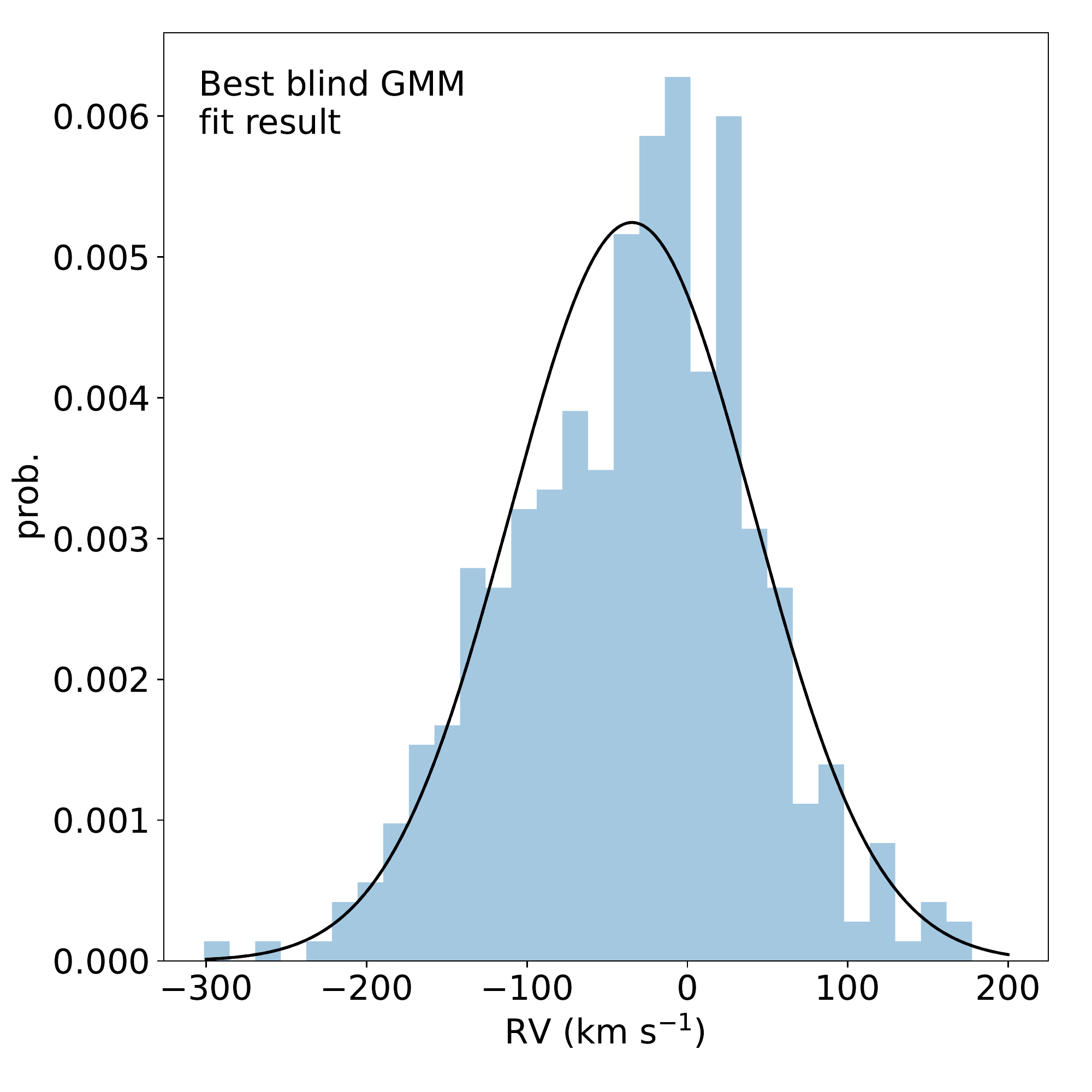}
\includegraphics[width=0.8\columnwidth]{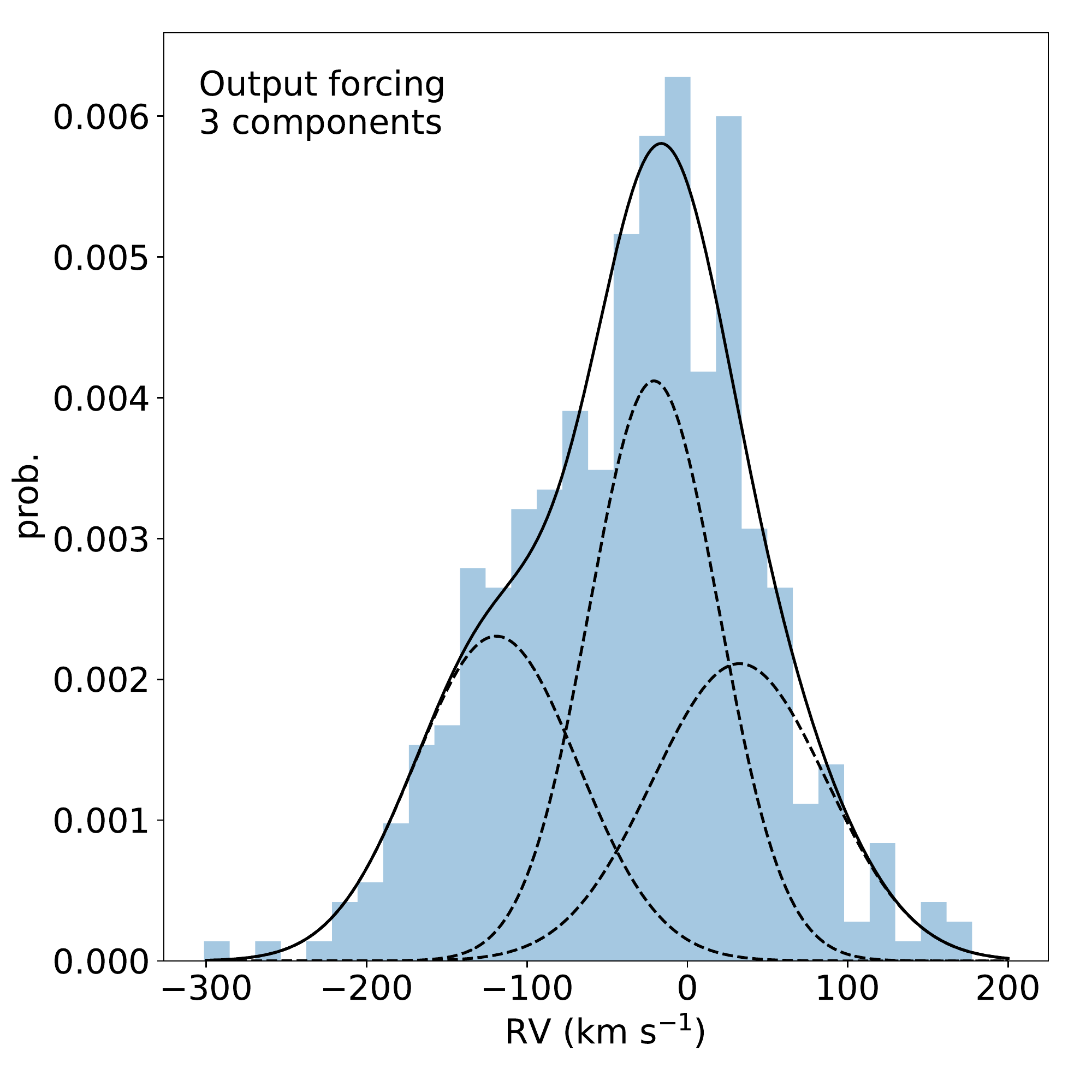}
\caption{GMM fitting results for a simulated distribution of 450 stars with three RV components with the mean and dispersion adopted for bulge and disc, and that derived for FSR\,1776. The top panel shows the input Gaussians that represent the bulge, disc, and cluster, used to generate the histogram in blue. The middle panel shows the best result with (RV,$\sigma$) $= (-34,~54)~{\rm km\, s^{-1}}$. The bottom panel shows the result forcing the GMM to find three components: (RV,~$\sigma$,~fraction) $= (-21,~29,~42\%),~(33,~38,~29\%),~(-119,~36,~29\%)~{\rm km\, s^{-1}}$.}
\label{fig:gmm}
\end{figure}

We have tested the ability of a GMM analysis to recover input RV populations from a simulated mixture. We generated a RVD made up of three components representing bulge, disc, and cluster with (RV,~$\sigma$,~fraction) $=(-50,~90,~53\%),~(0,~40,~35\%),~(-110,~32,~12\%)~{\rm km\, s^{-1}}$, in a total of 450 stars, to resemble the conditions of the MUSE RVD analysed in this work. The blind fit resulted best with a single component at (RV,$\sigma$) $= (-34,~54)~{\rm km\, s^{-1}}$. If we forced the GMM fit to find three components, the results were: (RV,~$\sigma$,~fraction) $= (-21,~29,~42\%),~(33,~38,~29\%),~(-119,~36,~29\%)~{\rm km\, s^{-1}}$. In conclusion, the GMM would not be able to blindly find the cluster RV signature with the MUSE data, as it is evident in Fig. \ref{fig:gmm}. Consequently, the use of prior information on bulge and disc kinematics is indeed necessary to look for the cluster in the residuals, as we have successfully performed in Sect. \ref{sec:RV}.

%
\section{Gaia radial velocities}
\label{app:gaiaRV}

\begin{figure}[htb]
\centering
\includegraphics[width=\columnwidth]{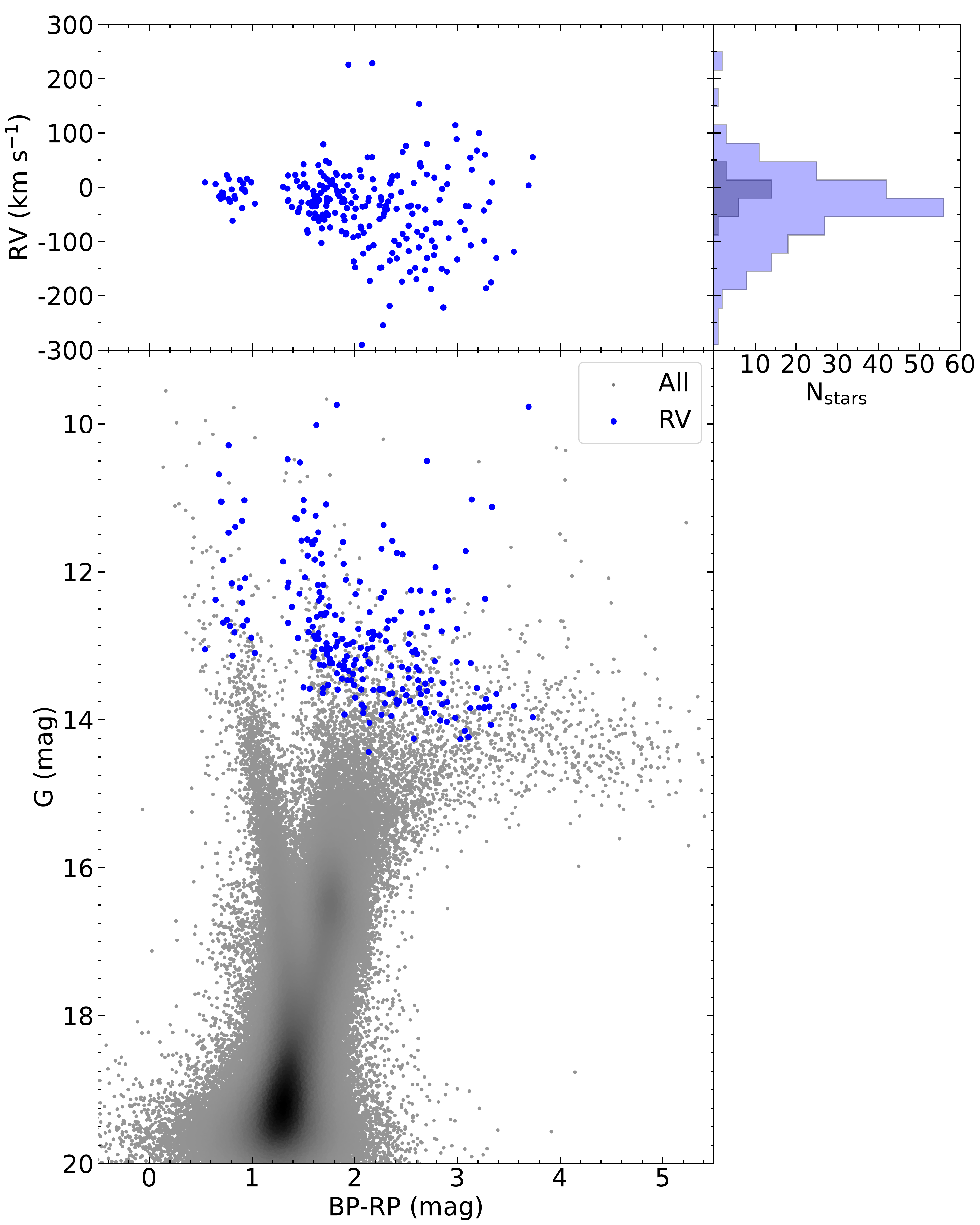}
\caption{{\it Bottom:} Colour-magnitude diagram for Gaia DR2 within $20\arcmin$ around FSR\,1776. The brightest stars ($G \lesssim 14$~mag) that have RV information are shown in blue. {\it Top left:} Colour vs. RV, where it is clear that there is a group of low RV and colour dispersion with BP-RP<1.2 and another with larger RV and colour dispersion with BP-RP >1.2. {\it Top right:} Histogram of RV for stars with BP-RP<1.2 and BP-RP >1.2 representing the younger disc (grey) and the older bulge (blue) components. It becomes evident that the assumptions for disc and bulge RV and dispersion in Sect. \ref{sec:RV} are consistent with Gaia data.}
\label{fig:gaiaRV}
\end{figure}

We have checked the available RVs from Gaia DR2 within $20\arcmin$ around the FSR\,1776 centre, which are the most up-to-date RVs from the Gaia mission. Unfortunately, there are RVs only for stars brighter than $G \lesssim 14$~mag, which is above the brightest stars from FSR\,1776 (see Fig. \ref{fig:gaiaRV}). Nevertheless, we show that the blue bright stars have RV$\sim0~{\rm km\, s^{-1}}$, which accounts for the disc component. The red bright stars span a range in colour, that can be seen as a span of metallicities and reveal an over-density in RV with peak around RV$\sim-50~{\rm km\, s^{-1}}$ resembling the bulge component. In conclusion, although Gaia RV does not reach cluster stars, it does endorse the assumed RVs for disc and bulge, as discussed in Section \ref{sec:RV}.

\end{document}